\documentclass[draft]{agujournal2019}
\usepackage{url} 
\usepackage{lineno}
\usepackage[inline]{trackchanges} 
\usepackage{soul}
\draftfalse

\journalname{JGR: Space Physics}

\begin{document}

%
%


\title{Mirror mode storms observed by Solar Orbiter}

%
%




\authors{A. P. Dimmock\affil{1}, E. Yordanova\affil{1}, D. B. Graham\affil{1}, Yu. V. Khotyaintsev\affil{1}, X. Blanco-Cano\affil{2}, P. Kajdič\affil{2}, T. Karlsson \affil{3}, A. Fedorov \affil{4}, C. J. Owen\affil{5}, E. A. L. E. Werner\affil{1}, A. Johlander\affil{1}}


\affiliation{1}{Swedish Institute of Space Physics, Uppsala, Sweden}
\affiliation{2}{Departamento de Ciencias Espaciales, Instituto de Geofísica, Universidad Nacional Autónoma de México, Ciudad Universitaria, Ciudad de México, Mexico}
\affiliation{3}{Division of Space and Plasma Physics, School of Electrical Engineering and Computer Science, KTH Royal Institute of Technology, Stockholm, Sweden}
\affiliation{4}{IRAP UPS CNRS, Toulouse, France}
\affiliation{5}{Mullard Space Science Laboratory, University College London, UK}




\correspondingauthor{Andrew P. Dimmock}{andrew.dimmock@irfu.se}




\begin{keypoints}
\item Mirror mode storms predominantly occurred during slow solar wind
\item Heliospheric plasma sheet crossings were effective at setting up MM unstable conditions
\item Spatial scales of mirror mode structures approached and were smaller than ion-scales
\end{keypoints}

%
%

%
%


\begin{abstract}
Mirror modes are ubiquitous in space plasma and grow from pressure anisotropy. Together with other instabilities, they play a fundamental role in constraining the free energy contained in the plasma. This study focuses on mirror modes observed in the solar wind by Solar Orbiter for heliocentric distances between 0.5 and 1 AU. Typically, mirror modes have timescales from several to tens of seconds and are considered quasi-MHD structures. In the solar wind, they also generally appear as isolated structures. However, in certain conditions, prolonged and bursty trains of higher frequency mirror modes are measured, which have been labeled previously as mirror mode storms. At present, only a handful of existing studies have focused on mirror mode storms, meaning that many open questions remain. In this study, Solar Orbiter has been used to investigate several key aspects of mirror mode storms: their dependence on heliocentric distance, association with local plasma properties, temporal/spatial scale, amplitude, and connections with larger-scale solar wind transients. The main results are that mirror mode storms often approach local ion scales and can no longer be treated as quasi-MHD, thus breaking the commonly used long-wavelength assumption. They are typically observed close to current sheets and downstream of interplanetary shocks. The events were observed during slow solar wind speeds and there was a tendency for higher occurrence closer to the Sun. The occurrence is low, so they do not play a fundamental role in regulating ambient solar wind but may play a larger role inside transients.
\end{abstract}

\section*{Plain Language Summary}
Plasma strives to be in equilibrium with little to no free energy. However, this is often not the case, especially in close proximity to complex structures such as shock waves and interplanetary coronal mass ejections. The latter is an eruption of plasma from the Sun that propagates outward into the solar system. In the presence of some free energy, instabilities will arise to remove it, one example is the mirror mode instability. Instabilities such as these are of extremely high importance to plasma physics as they act as a feedback mechanism to the plasma. Nevertheless, there are many open questions regarding the mirror mode instability, especially when their properties are different from the most common scenarios. Typically, mirror modes in the solar wind appear as dips that are isolated structures. However, this paper investigates mirror modes when they appear as sudden bursts of magnetic peaks and dips and typically have smaller temporal scales. These kinds of mirror modes have been called mirror mode storms. This study aims to address at what distances from the Sun they arise, what types of solar wind structures they are associated with, quantify their physical properties, and understand what local plasma conditions are important.

%
%

%


%
%
%
%

\section{Introduction}

Mirror modes (MMs) are fundamental plasma phenomena that are universal across a diverse set of space plasma environments \cite{Tsurutani82, Neubauer93, Joy06, Genot08, Genot09,soucek2008,balikhin2009,soucek2015}. Analogous to other plasma instabilities, MMs are essential to understanding both the global and local kinetic behavior of plasma as they are a natural feedback mechanism that drives the plasma towards marginal stability. Through theory, MMs were first predicted \cite{chandrasekhar1958, hasegawa1969} until the observational evidence arrived soon after \cite{kaufmann1971}. What ensued was a multitude of MM observations \cite{tsurutani1982, neubauer1993, sahraoui2004, joy2006, volwerk2008,genot2009,soucek2008,balikhin2009,soucek2015,osmane2015,dimmock2015,volwerk2016, ala-lahti2018,karlsson2021} in regions such as the solar wind, planetary magnetosheaths, Interplanetary Coronal Mass Ejections (ICMEs), and around comets. Furthermore, MMs have also been studied in the context of local and global numerical simulations \cite{hoilijoki2016,ahmadi2017}.

Although they are commonly treated from a quasi magnetohydrodynamic (MHD) perspective, they are kinetic structures by nature. They have zero phase velocity in the plasma rest frame, appear as sharp peaks or dips in the magnetic field that are anti-correlated with density, and are linearly polarized. MMs grow when there is sufficient free energy from the ion pressure anisotropy ($Pi_A = P_{\perp i}/P_{\parallel i} > 1$) and the plasma $\beta_i$ is sufficiently high. The perpendicular pressure constructs local magnetic mirror configurations analogous to a magnetic bottle. Particles undergo mirror motion between the so-called bottlenecks, which results in the anti-correlation between the magnetic field and particle density when traversed by a spacecraft. \citeA{hasegawa1969} derived a convenient threshold to describe mirror unstable plasma ($T_{\perp i}/T_{\parallel i} > 1 + 1/\beta_{\perp i}$) based on a bi-Maxwellian cold electron fluid approximation, and thus is valid when $T_{\perp e} \sim T_{\parallel e} \ll T_{\parallel i}$. This threshold is based on a kinetic theory at the long-wavelength limit (see eqs 2-4 in \citeA{hasegawa1969}). Thus, although a kinetic approach is used, its use is applicable when spatial wavelengths are much greater than the ion gyroradius (i.e. $L_{mm} \gg \rho_p$), where $L_{mm}$ is the spatial scale of one MM structure and $\rho_p$ is the proton gyroradius. Thus, MMs are often referred to as quasi-MHD. The MM threshold establishes that the local $\beta_i=2\mu_0nk_B T_i/B^2$ and $T_{\perp i}/T_{\parallel i}$ are necessary to quantify the degree of stability of plasma to MMs. Moreover, for $T_{\perp i}>T_{\parallel i}$ conditions, the MM instability competes with the Alfv\'en ion cyclotron (AIC) instability that dominates at lower values of plasma $\beta_i$ \cite{gary1992}. For completeness, it is also worth mentioning the firehose instability, which grows when $T_{\parallel i}>T_{\perp i}$, implying it is mutually exclusive with the MM and AIC instabilities. Nevertheless, the content of this paper will focus explicitly on MMs.

MMs are frequently observed in planetary magnetosheaths as the shocked solar wind plasma provides favourable conditions ($\beta_i>1$, $T_{\perp i}>T_{\parallel i}$) for MM growth \cite{volwerk2008,soucek2008,genot2009,dimmock2015}. The readily available high-cadence measurements from missions such as Cluster, THEMIS, and MMS have been used to characterize and study MMs in the Earth's magnetosheath \cite{genot2009,soucek2015,dimmock2015}. In general, MMs in the Earth's magnetosheath appear in the form of continuous trains of peaks or dips \cite{soucek2008,genot2009,dimmock2015} with average temporal periods $\sim 13$ s \cite{soucek2008}. Considering the average flow speeds in the magnetosheath \cite{dimmock2013}, then the spatial extent of these structures approaches fluid scales. The MM ``peakness" is typically identified based on the skewness of the probability distribution of the magnetic field; where negative values suggest the existence of dips and vice versa in the case of peaks. The occurrence of peaks or dips is understood to be related to the degree of instability of the plasma \cite{soucek2008,genot2009,dimmock2015}. Peaks are associated with MM unstable plasma whereas dips appear around or below marginal stability. Together, the MM and AIC instabilities, both a function of $\beta_i$, put an upper bound on the ion temperature anisotropy that produces a clear anti-correlation between $T_{\perp i}/T_{\parallel i}$ and $\beta_i$ \cite{gary1994,fuselier1994}. Regardless, MMs can also be excited by electron anisotropies. \citeA{yao2019} presented a case study of the electron MM (scales below proton gyroradius) in the Earth's magnetosheath corresponding to the condition $T_{\perp e} /T_{\parallel e} > 1 + 1/\beta_{\perp e}$. In this event, there was no ion temperature anisotropy but a clear electron temperature anisotropy was present that was in anti-correlation with the electron pressure. These structures appeared as trains of dips. Kinetic scale magnetic dips have also been reported in the magnetosheath as more isolated structures \cite{yao2019b}

MMs are also observed inside the sheath regions of interplanetary coronal mass ejections (ICMEs), occurring in around 70\% of the cases at 1 AU behind the leading IP shock \cite{ala-lahti2018}. Despite this high occurrence rate, studies that have focused explicitly on MMs inside ICME sheaths are uncommon (e.g. \citeA{liu2006,ala-lahti2018}). The recent statistical study by \citeA{ala-lahti2018} estimated the occurrence and physical properties of MMs measured inside 96 ICME sheaths at 1 AU using the Wind spacecraft. The MMs displayed an average temporal period between 11.6 s-13.7 s depending on if they were part of a MM train or isolated structures; the general temporal width varied from around 6 s to over 40 s. Hence, the spatial scales should be on the order of thousands of km, much larger than the hundreds of km expected from the ion gyroradius. Thus, the long-wavelength approximation should be valid. There was also large variability in the wave amplitudes (1 nT-14 nT). According to the statistical distribution from the events considered, the structures had amplitudes of approximately 3 nT and 96\% of the time were dips. Although MMs inside ICME sheaths can appear as trains, they are not as tightly packed and successive as those seen in the Earth's magnetosheath.

Structures in the solar wind called magnetic holes have been reported for decades \cite{turner1977,Winterhalter1994,Xiao2010,volwerk2020,karlsson2021}, and resemble MM structures. These generally differ from the MM train-like structures seen in magnetosheaths since they are especially more isolated and manifest at larger temporal and spatial scales. Their scale sizes range from several seconds to minutes (see \citeA{karlsson2021} and references therein). At 1 AU, the occurrence rates are between 2.4 - 3.4 holes per day, for those that are linear with no field rotations before and after the hole \cite{pokhotelov2002}. There are striking resemblances between linear magnetic holes and MMs, such as the pressure balance, linear polarization, and tendency to occur in regions unstable to the MM instability criteria \cite{Tsurutani2010}. However, it has also been shown that magnetic holes can occur in mirror stable plasma \cite{stevens2007}, so open questions still remain. Innately, it has been proposed that magnetic holes could be remnants of the MM instability in localized regions \cite{Winterhalter1994}. Nevertheless, magnetic holes are frequently observed in the solar wind across varied heliocentric distances. Yet, in some cases, MMs materialize in the solar wind with properties that are significantly different from magnetic holes.

MM structures also occur in the solar wind in the form of prolonged trains, which are remarkably similar to those reported in planetary magnetosheath regions \cite{russell2009, enriquez_rivera2013}. They maintain low amplitudes ($\sim 1$ nT) and manifest as peaks or dips. These events have been designated mirror mode storms (MM storms) \cite{russell2009} but the literature is scarce; to our knowledge, just a few studies have been published (e.g. \citeA{russell2009,enriquez_rivera2013}) to date. Using STEREO measurements, \citeA{enriquez_rivera2013} reported on MM storms by characterizing 15 events and then conducting a kinetic dispersion analysis. Most of their events were observed for stream interaction regions (SIRs) and only one was associated with the ambient solar wind. Interestingly, the authors note that alpha particle density also increased for most of their MM storm events. Nevertheless, in regions of high $\beta_i$, the ion temperature anisotropy needed for the plasma to become mirror mode unstable diminishes, and SIRs can offer the ideal conditions. Interestingly, the kinetic analysis suggested that ion cyclotron waves should also be generated for similar conditions but were not observed. They suggested that the differing phase velocities may be responsible for the absence of concurrent observations. It has been understood for some time that particularly in planetary magnetosheaths the MM and ion cyclotron instabilities compete depending on the local $\beta_i$ \cite{soucek2015}.

The current study utilizes data from Solar Orbiter (SolO) to study MM storms at heliospheric distances between 0.5-1 AU. The goal and motivation for this study were to contribute to filling this gap and shed light on some unresolved questions. This was achieved by employing the novel SolO observations to investigate characteristics such as physical properties (e.g. amplitude, frequency, peaks/dips, spatial scale), dependence on local plasma conditions, and connection with solar wind structures (e.g. SIRs and shocks), and their occurrence across heliocentric distances. The study firstly analyzed several case studies in detail before conducting an automated search for events. This produced 25 events that were used to investigate the occurrence rate, dependence on solar wind conditions, and location in the inner-heliosphere.

\section{Data \& Instrumentation}
SolO \cite{muller2020} measurements collected between 2020-04-15 and 2021-08-31 are used to conduct this investigation. The fluxgate magnetometer instrument (MAG) \cite{horbury2020} provides full 3D magnetic field vectors and is used to characterize the magnetic field properties of the large-scale structure and MM waves. The magnetic field data are also used to automatically detect MMs later in the paper. The radio and plasma wave experiment (RPW) \cite{maksimovic2020} measures the probe-to-spacecraft potential ($ScPot$), which can be calibrated to estimate the local electron density ($N_e$) \cite{khotyaintsev2021}. Hereafter, $N_e$ refers to electron density from $ScPot$ and is not calculated from moments of velocity distributions. This high temporal resolution is needed since MM storms are typically around 0.5-1 Hz in contrast to solar wind magnetic holes and typical solar wind mirror modes that are above several seconds. The solar wind analyzer (SWA) instrument \cite{owen2020}, particularly the proton alpha sensor (SWA-PAS), is then employed to provide ion velocity distribution functions (VDFs) and ion moments. Note that ground-based moments are determined from the proton peak in the SWA-PAS VDFs. However, the proton peak cannot always be easily distinguished and thus alphas may sometimes affect the moment calculations. However, there is no dedicated flag to indicate when the proton and alpha peaks are well resolved, so this study does not consider such effects. The electron analyzer system (SWA-EAS) was also used to obtain electron pitch angle distributions. Also note that the rtn (radial, tangential, normal) coordinate system is used unless stated otherwise. The CDF file versions used to analyze individual MM events (and compute solar wind statistics) were V1-V3 (V1-V5) for MAG and V3 (V3) for SWA-PAS. Measurements from the Magnetospheric Multiscale Mission (MMS) were also employed from the fluxgate magnetometer \cite{russell2016}, Fast Plasma Investigation-Dual Ion Spectrometer (FPI-DIS) \cite{pollock2016}. OMNI data was used to infer ion temperature for the MMS event since FPI-DIS was not intended to measure the solar wind.

\section{Case studies}
\subsection{Event 1: 2021-07-19}
\begin{figure}
\noindent\includegraphics[width=0.75\textwidth]{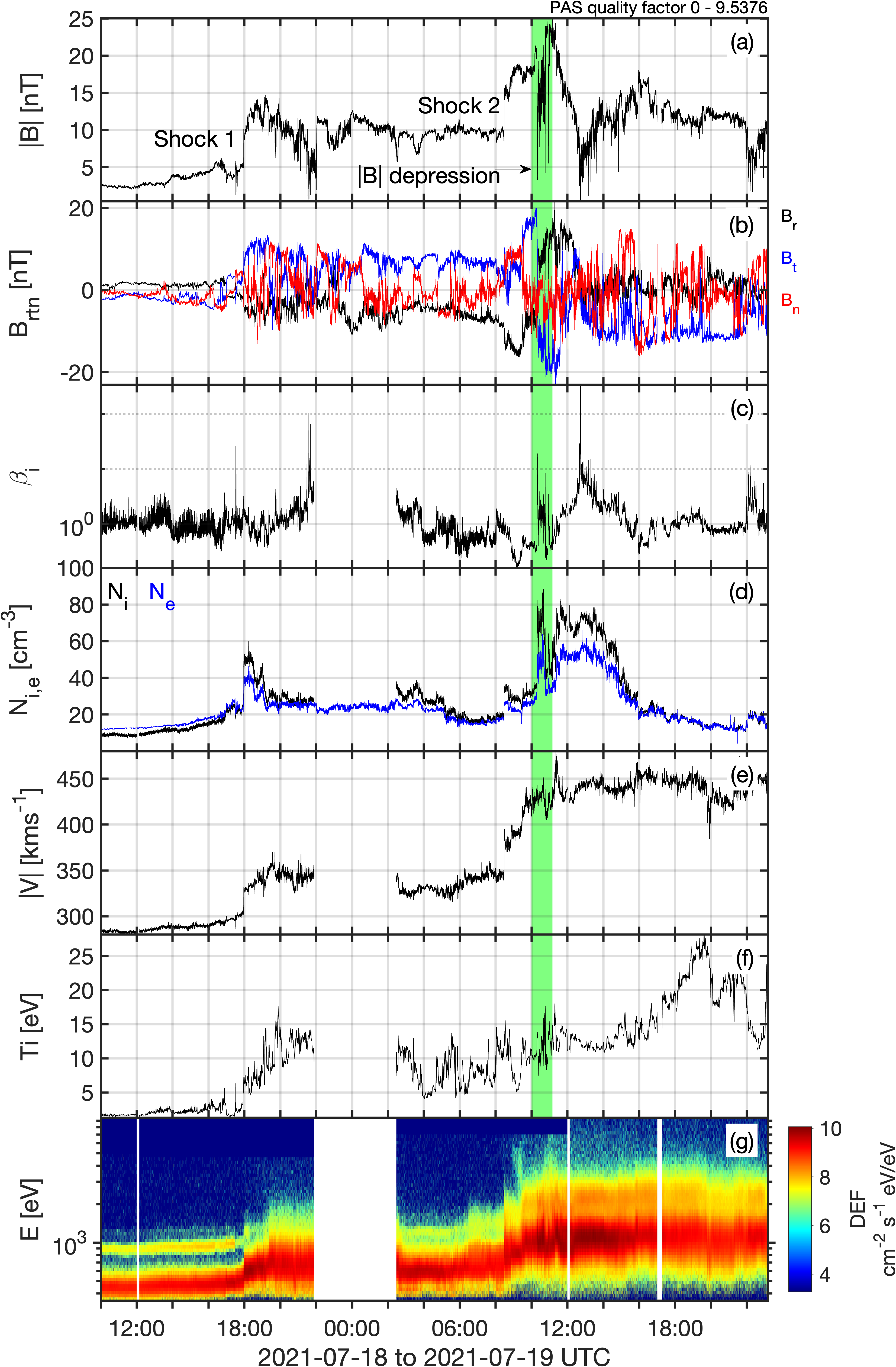}
\caption{Overview of the event for the interval on days 2021-07-18 and 2021-07-19. Panels (a-f) correspond to $|\mathbf{B}|$, $B_{rtn}$, $\beta_i$, $N_i$, $|\mathbf{V_i}|$, and $T_i$, respectively. The bottom panel (g) shows the omnidirectional DEF. The area highlighted in green shows the region of interest, which contained abundant MM structures.}
\label{fig1}
\end{figure}

Plotted in Figure \ref{fig1} are SolO measurements collected between 10:00 on 2021-07-18 and 23:10 on 2021-07-19 when the spacecraft was 0.84 AU from the Sun. Panels (a \& b) show the magnetic field while the remaining panels (c-g) correspond to $\beta_i$, $N_{i,e}$, $|\mathbf{V_i}|$, $T_i$ and omnidirectional differential energy flux (DEF). 
Near 18:00 on 2021-07-18, SolO measured a fast forward shock according to the concurrent increase of $|\mathbf{B}|$, $N_i$, and $|\mathbf{V_i}|$ in panels (a, d, \& e). Later, about 08:30 on 2021-07-19, another fast forward shock was measured according to comparable signatures in $|\mathbf{B}|$, $N_i$, and $|\mathbf{V_i}|$. Before this event, the solar wind speed was slow, below 300 kms$^{-1}$ but then increased at the shock crossings to eventually 450 kms$^{-1}$ at the end of the interval. $N_i$ was highly varying over the entire event, rising to around 60 cm$^{-3}$ at the first shock but afterward increasing further to over 80 cm$^{-3}$. The ions are also heated at both shock crossings as shown by the sudden step increases and broadening of the omnidirectional spectra in panel (g). Also meaningful are instances of unusually small $|\mathbf{B}|$ to almost zero that creates large values in $\beta_i$, which is discussed later. There are also substantial rotations of the magnetic field in panel (b) signifying complex structures such as embedded flux ropes and/or current sheets. The large-scale features and double shock crossings are consistent with the passage of an SIR. The structured SIR region results from the interaction between a slow wind stream and a fast wind stream, where shock waves separate the unperturbed solar wind from the shocked slow wind compressed by the incoming fast stream. Then another shock separates the slowed down and compressed fast wind and the trailing undisturbed fast stream. This picture corresponds to a two-stage increase in the density, magnetic field magnitude and plasma temperature, along with the typical transition from low to high plasma flow speed \cite{Richardson2018}.

\begin{figure}
\noindent\includegraphics[width=0.75\textwidth]{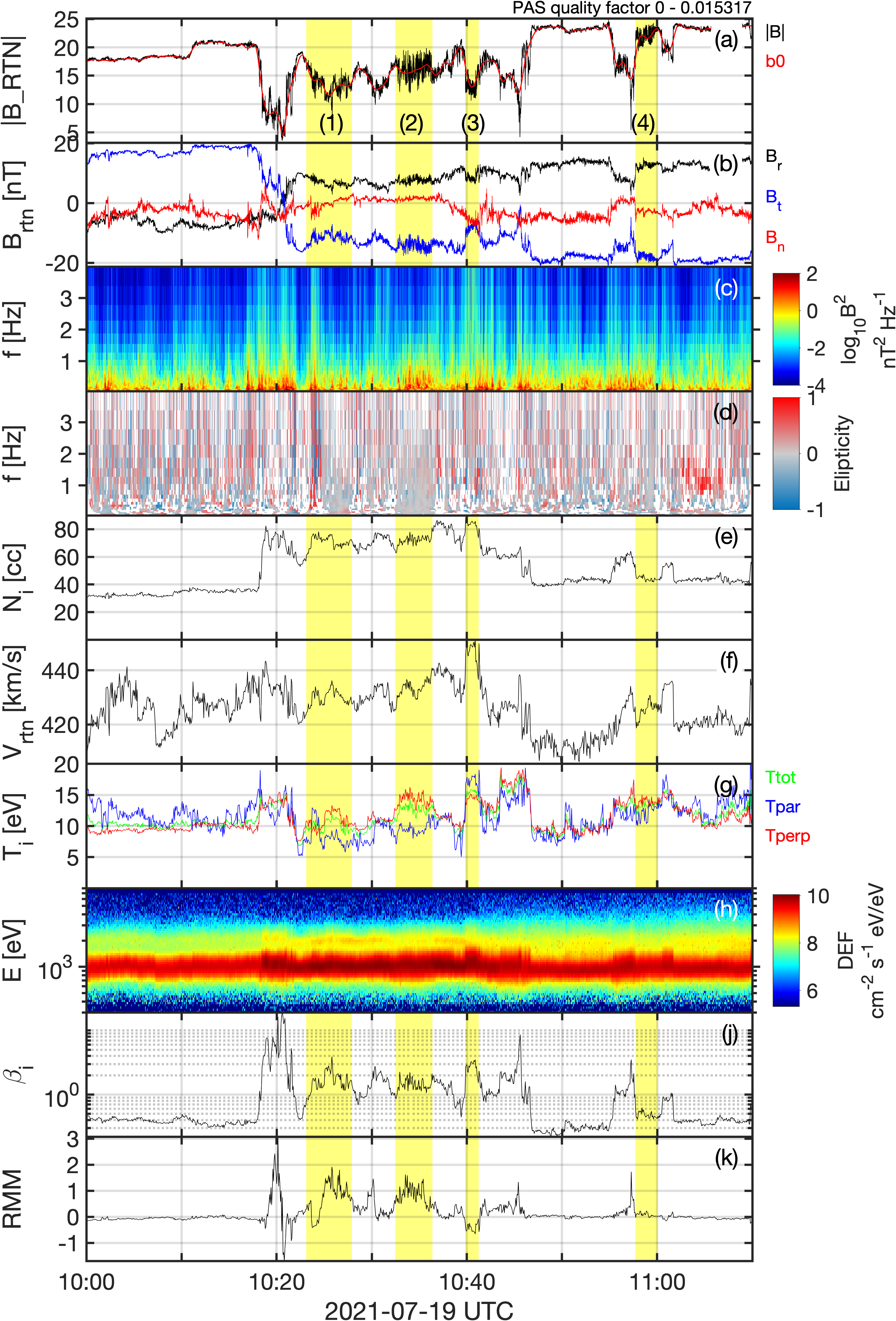}
\caption{MMs observed on 2021-07-19. Plotted in panels (a \& b) are $|\mathbf{B}|$ and $B_{rtn}$, a wavelet spectrogram of $\mathbf{B}$ is shown in panel (c), and the ellipticity of the magnetic field is shown in panel (d). Panels (e-k) depict $N_i$, $|\mathbf{V_i}|$, $T_i$, DEF, $\beta_i$, and $RMM$, respectively. Regions that are highlighted in yellow correspond to localized reductions in ellipticity and the manifestation of MM structures since they should have zero ellipticity.}
\label{fig2}
\end{figure}

\begin{figure}
\noindent\includegraphics[width=\textwidth]{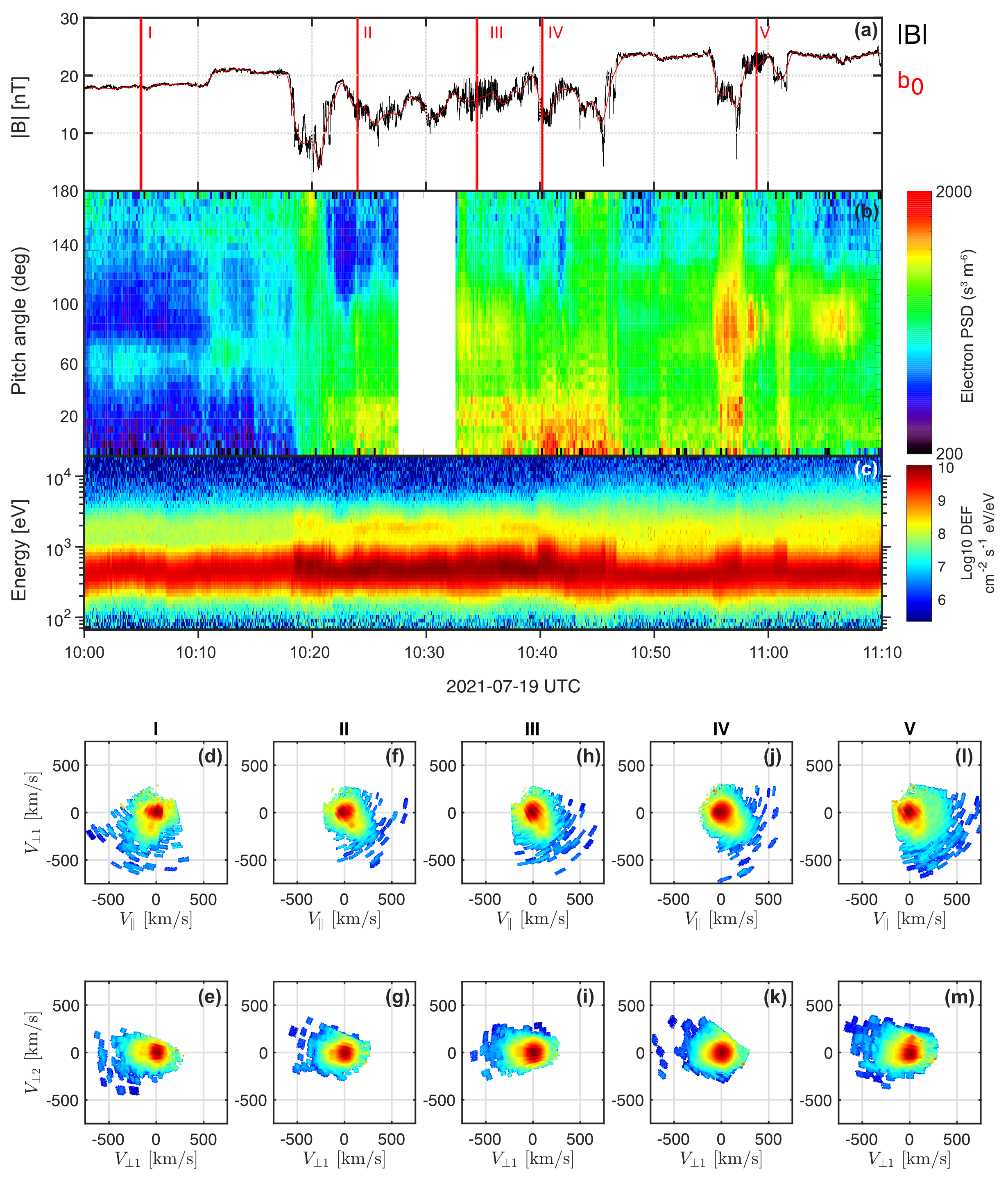}
\caption{Evolution of ion VDFs and electron pitch angle. Panels (a c) show $|\mathbf{B}|$, ion DEF, and electron pitch angle phase space density, respectively. Panels (d-m) are VDFs at the time instances marked by I-V in panel (a). The top row is a 2D reduced distribution in the $V_\parallel$-$V_{\perp1}$ plane whereas the bottom row is the $V_{\perp1}$-$V_{\perp2}$ plane.}
\label{fig3}
\end{figure}

In addition to the large-scale variations of the magnetic field and plasma parameters seen in Figure \ref{fig1}, smaller-scale waves and structures were also observed in concert. The large-scale variations appear inherently connected to these smaller scales as they are responsible for significantly modifying the local conditions that favor the growth of waves and instabilities. Of distinct interest to this study is the area highlighted in green around 10:00-12:00 UT on 2021-07-19. This interval contains a significant magnetic depression and a polarity reversal of the radial and tangential magnetic field components. The density also increases by a factor of two from 40 cm$^{-3}$ to over 80 cm$^{-3}$, and although no substantial changes in velocity occurred, there were variations in the temperature moments and intricate features in the DEF spectra. The interpretation is that perhaps the spacecraft was crossing the heliospheric plasma sheet (HPS). In \citeA{Simunac2012} such crossings were identified by magnetic field polarity changes, increased plasma density, a local decrease in the alpha particle-to-proton number density, and a local increase in the ion density. In our case, electron pitch-angle distributions support a crossing of the heliospheric current sheet HCS at $\sim$ 10:18 UTC, where the electron distributions change from anti-parallel to parallel (see Figure \ref{fig3} later panel b). Interestingly, inside this potential HPS encounter, numerous bursts of linearly polarized structures were recorded. A more thorough analysis of these structures is presented in Figure \ref{fig2}.

The HPS marked by the green highlighted region in Figure \ref{fig1} is shown in Figure \ref{fig2}. A wavelet spectrogram of $\mathbf{B}$ is added in panel (c) and the ellipticity of the magnetic field \cite{santolik2003} is plotted in panel (d). The ellipticity of $\pm$1 corresponds to right-/left-handed circular polarization, and 0 to linear polarization. Ion temperature in magnetic field-aligned coordinates is located in panel (g). $\beta_i$ is plotted in panel (j) and the MM instability criterion $RMM$ is plotted in the bottom panel. Note here (and in other figures), $\beta_i$ is plotted but $\beta_{\perp i}$ is used in the calculation for $RMM$. The quantity $RMM$ \cite{soucek2008} provides a measure of the variation from stability and is calculated  as follows:
  \begin{linenomath*}
 \begin{equation}
 RMM = \beta_{\perp i} \left(\frac{T_{\perp i}}{T_{\parallel i}}-1 \right).
 \label{eq:R}
 \end{equation}
  \end{linenomath*}
Here, and throughout the paper, the background magnetic field is determined using a low-pass filter with a cut-off frequency of 0.01 Hz. Instability to MMs corresponds to $RMM>1$ but mirror modes are also shown to appear when $RMM<1$. In general, when $RMM>1$ (unstable plasma) MMs are peaks but appear as dips when $RMM<1$. Equation \ref{eq:R} implies that MMs will grow when a temperature anisotropy is present. In reality, the situation is more complex since MMs compete with the ion cyclotron instability depending on the ion plasma $\beta$. In general, the ion cyclotron instability will dominate for lower $\beta$ \cite{soucek2015}.

During the HPS encounter, there was a pronounced increase in the spectral power of $\mathbf{B}$ as seen in Figure \ref{fig2}c, suggesting the presence of waves and/or enhanced turbulence. More information is provided in panel (d) by calculating the magnetic field ellipticity when the degree of polarization was above 0.8. This unveiled multiple bursts of linearly polarized structures, which were highlighted in yellow (1-4). As said previously, between 10:20 and 11:00, the ion density increased significantly ($\sim$ 40-80 cm$^{-3}$) but what was evident over this timescale is the complex behavior of the ion temperature anisotropy. In general, $T_{\perp i}>T_{\parallel i}$ over this interval, and the DEF intensity increased close to 1 keV while there appeared to be an increase in alpha particle density. Evidence of an alpha particle density increase is seen from the enhancement in DEF above the main population (i.e. $>$1 keV). Upon closer inspection, the yellow highlighted intervals were consistent with the characteristics of  MM structures. According to panels (c \& d), the time period of these structures was around and slightly below 1 second, corresponding to wavelengths roughly 400 km assuming zero phase speed in the plasma rest frame. The ion gyro-radius ($\rho_p$) within this interval was around 32 km implying that $L_{mm}\sim 13 \rho_p$, where $L_{mm}$ is the spatial scale of an individual MM structure. Intervals 1 and 2 display enhancements in $RMM$ in panel (k), as expected from the concurrent increase in $T_{\perp i}/T_{\parallel i}$ and $\beta_i$ from panels (g \& j). Interval 3 was MM stable ($RMM<1$) due to $T_{\parallel i}>T_{\perp i 1}$,and therefore surprising that MMs were so prevalent. On the other hand, since MMs are convected with the plasma flow, such in situ conditions may not match the plasma parameters at the moment when the MM structures were generated; this is a plausible scenario considering the variations of $RMM$ over this brief interval. The final interval was intriguing as $RMM\sim0$, which will be discussed later. 

In Figure \ref{fig3}, VDFs were presented at five instances that were marked by the vertical red lines in panel (a) by roman numerals I-V. These were chosen to provide an overview of the changes in the VDFs across the event. The VDFs are shown in two planes according to $\parallel-\perp1$ and $\perp1 - \perp2$, which is derived with respect to the background magnetic field. The DEF was shown for reference in panel (b) whereas the VDFs were placed below (d-m). The plotted VDFs are averages of five distributions, which correspond to 16 seconds and the width of the red lines. 
According to Figure \ref{fig3}, there were noticeable variations of the ion VDFs throughout this interval. For the majority of the event, the VDFs appeared moderately gyrotropic. There is a field rotation (reversal in $B_t$) around 10:18 and this effect is seen in the VDFs from column II as elongated shapes oblique to the local background magnetic field, which is present thereafter. This did result in temperature anisotropy and favorable MM growth between 10:20 and 10:40 also due to the high $\beta_i$. However, column IV did appear slightly different compared to II, III, and V, which coincided with a sharp change in $B_t$ and larger localized parallel temperature according to the moments, resulting in unfavorable conditions for MMs even though they are seen in $\mathbf{B}$. VDF IV seemed to evolve to more gyrotropic but the differences in the $V_{\perp1}$-$V_{\perp2}$ plane (k) constitute a few pixels and thus it is not possible to draw strong conclusions from this. Thus, for much of this interval, the MM growth condition seemed satisfied but still, MMs were present in locations where the VDFs and moments did not provide a clear explanation.

\subsection{Event 2: 31 May 2021}
\begin{figure}
\noindent\includegraphics[width=0.75\textwidth]{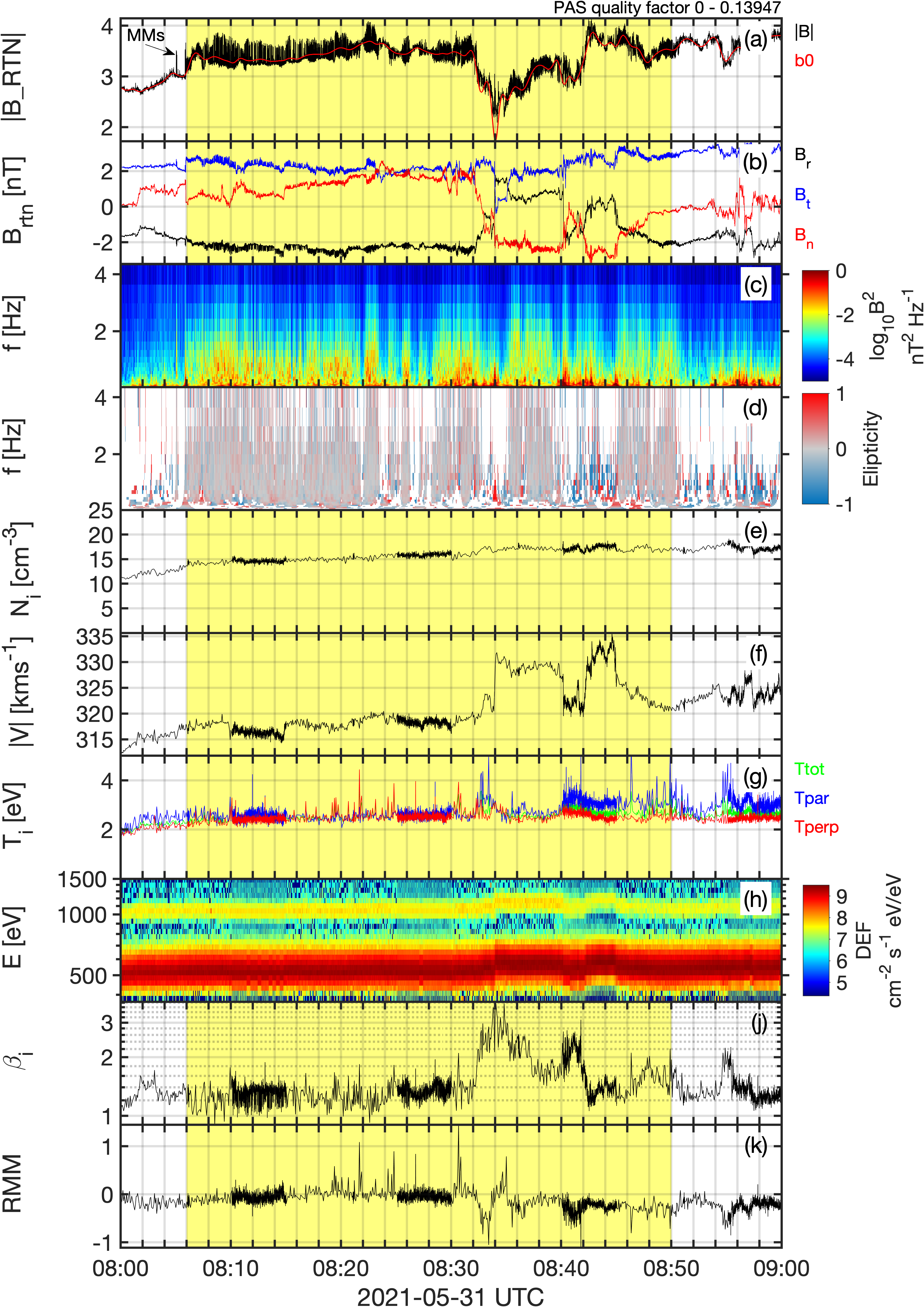}
\caption{MMs observed on 31 May 2021. Plotted in panels (a \& b) are $|\mathbf{B}|$ and $B_{rtn}$, a wavelet spectrogram of $\mathbf{B}$ is shown in panel (c), and the ellipticity of the magnetic field is shown in panel (d). Panels (e-k) depict $N_i$, $|\mathbf{V_i}|$, $T_i$, DEF, $\beta_i$, and $RMM$, respectively.}
\label{fig4}
\end{figure}

Presented in Figure \ref{fig4} is another period of intensive MM activity on 31 May 2021, when SolO was at a heliocentric distance of 0.95 AU. The layout of the panels is equivalent to those shown in Figure \ref{fig2}. Occasionally the SWA-PAS instrument would measure higher-cadence burst mode data for 5 minutes every 15 minutes, which is visible in the time series plot. The highlighted interval denotes the period of MM activity, which started after an increase in $|\mathbf{B}|$ at 08:08 due to abrupt changes in $B_n$ and $B_t$ (red and blue traces). Interestingly, it is worth remarking that isolated MM structures were also observed before this, such as the individual peak near 08:05, which has been marked in panel (a). Thus, the plasma was likely to be marginally MM unstable before the onset of the wave trains. Nevertheless, the advent of the MM trains coincided with a small decrease in $\beta_i$, a small increase in $N_i$, but no change in $V_i$ or $T_i$. Thus, this is not interpreted as a shock crossing. The MM instability threshold was also below zero for the majority of this interval.

Contrary to the event on 18 July 2021, these MMs appeared as extended trains of structures for over 40 minutes (highlighted in yellow) rather than shorter distinctive bursts of several minutes. The structures were linearly polarized, appeared as peaks, had periods of around 1 second, and amplitudes of roughly 0.5 nT. Based on the local plasma conditions, the spatial scale of these MMs was $L_{mm}\sim4.5 \rho_p$. Later in the interval, there was a polarity reversal of $B_n$ and $B_t$, and the MMs appeared suppressed. Yet, they began again soon afterward but were more bursty by nature, which could be reflective of the variable $\beta_i$. Confusingly, there was no significant temperature anisotropy over this period. As expected, $RMM$ remains predominantly below zero meaning the plasma was stable or marginally stable over this interval. These MMs also arose during a low energy slow solar wind stream, which exhibited a low ion temperature and moderate density. Evidently, the circumstances that led to this long train of MMs were different from the previous example.

\begin{figure}
\noindent\includegraphics[width=0.5\textwidth]{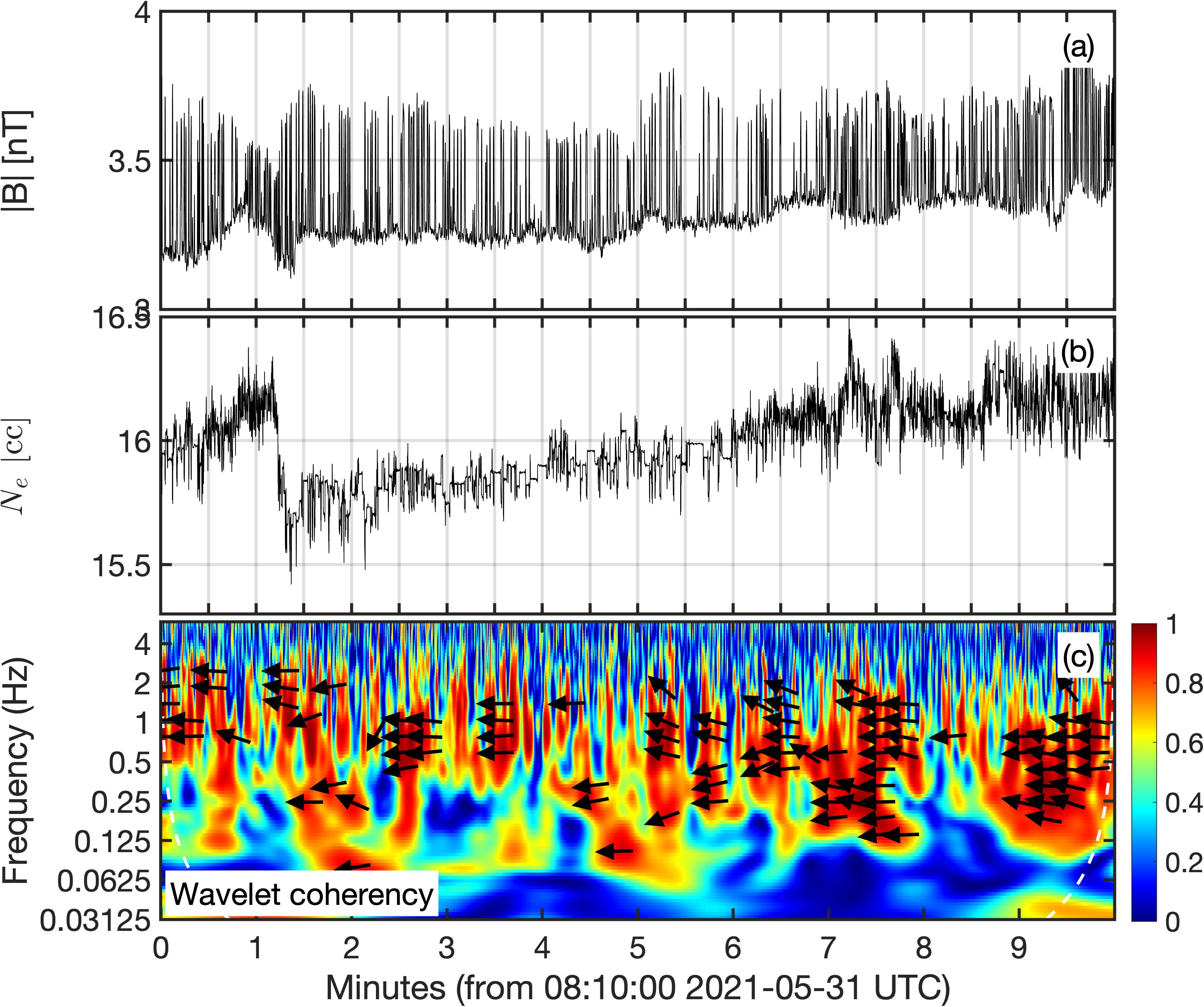}
\caption{Wavelet coherency between $|\mathbf{B}|$ and $N_e$. Panels (a \& b) show $|\mathbf{B}|$ and $N_e$ during MM activity on 31 May 2021 and the corresponding wavelet coherency spectra (c). The color in panel (c) depicts the coherency (0-1) whereas the arrows pointing left suggest anti-phase.}
\label{fig5}
\end{figure}

Figure \ref{fig5} reveals the MMs in more detail and the sharp peak structures are unmistakable from $|\mathbf B|$ in panel (a). Plotted in panel (b) is $N_e$ whereas panel (c) is a wavelet coherency spectra between $|\mathbf B|$ and $N_e$, which represents the coherency and phase between these quantities for the shown frequency range.
A fundamental attribute of MMs is the anti-correlation between $\mathbf B$ and density. The bulging local magnetic field induced by pressure anisotropy sets up bottle-like structures that create local magnetic mirror points. As a spacecraft transits through these structures then it will measure a time series of $|\mathbf B|$ and density that are anti-correlated \cite{soucek2008,dimmock2015}. The wavelet coherency confirmed this, demonstrating that the frequencies matching the MM time scales ($\sim 1 Hz$) displayed coherency values close to one and phase shifts around 180$^\circ$. The direction of the arrows denotes the phase such that pointing to the right (left) is in-phase (out of phase). Here, arrows are only plotted when the coherency exceeded 0.85 and the prevailing trend of these arrows is that they are pointing to the left and thus clearly demonstrate the anti-correlation over this interval. Note that the anti-correlation is only visible in $N_e$ since the cadence was sufficiently high compared to the ion moments. This anti-phase behavior was also observed for different events, but in some circumstances, it was not measurable due to the generally small amplitudes of MM storms.

\begin{figure}
\noindent\includegraphics[width=\textwidth]{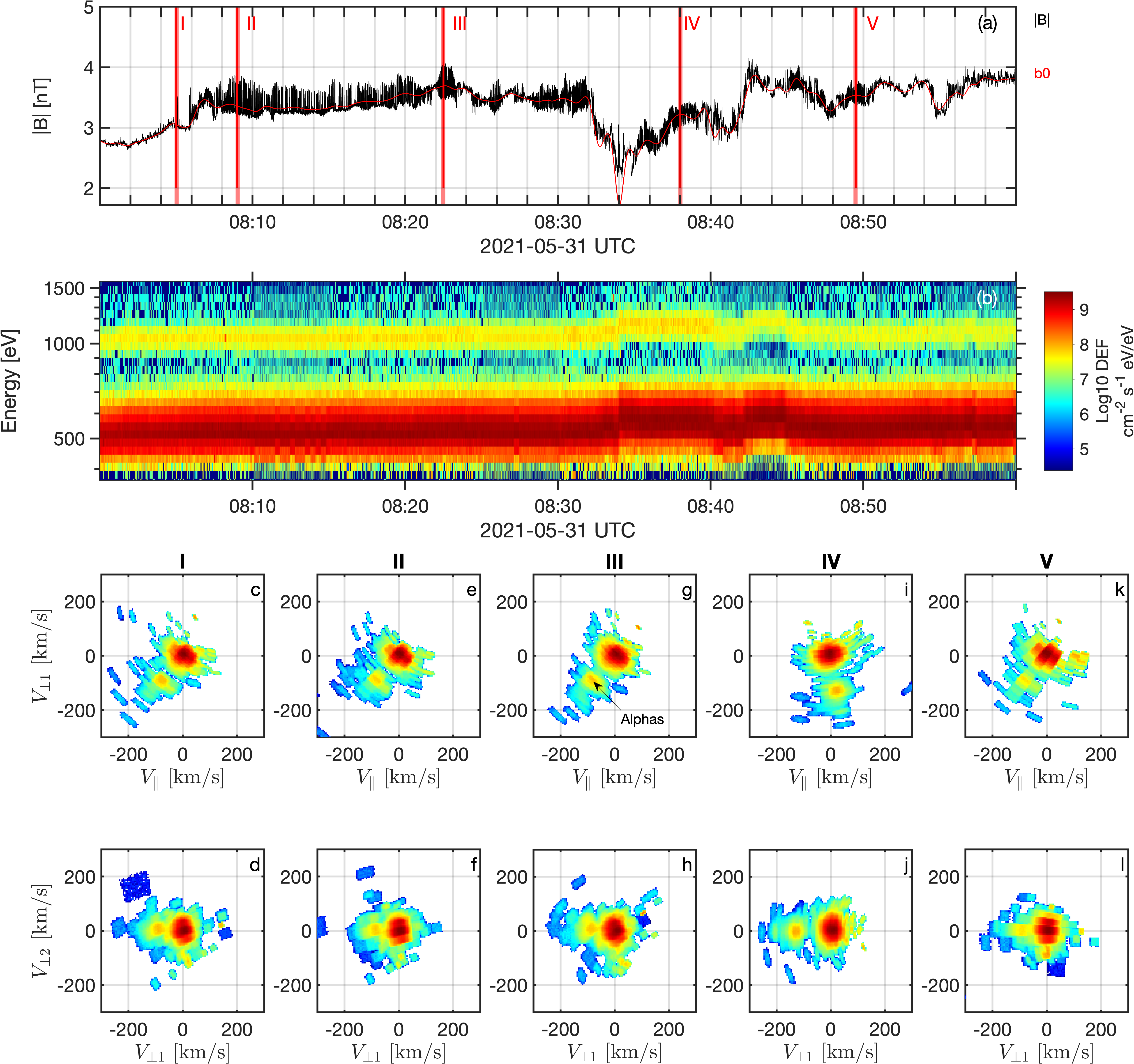}
\caption{Evolution of ion VDFs, Panels (a \& b) show $|\mathbf{B}|$ and and the DEF, respectively. Panels (c-l) are VDFs at the time instances marked by I-V in panel (a). The top row is a 2D reduced distribution in the $V_\parallel$-$V_{\perp1}$ plane whereas the bottom row is the $V_{\perp1}$-$V_{\perp2}$ plane.}
\label{fig6}
\end{figure}

Figure \ref{fig6} shows VDFs for the 31 May 2021 event at the times marked in panel (a) by the vertical red lines (I-V). Each VDF is an average of 5 VDFs, which is equivalent to the thickness of the vertical red lines. For reference, $|\mathbf{B}|$ and the DEF have been plotted in panels (a \& b).
The alpha particles are clear in panel (b) by the population around 1 keV above the solar wind at 500 eV. The VDFs measured by SWA-PAS measure both ions and alphas and in cases like this, the proton peak is well-defined. Thus, the moments should not be affected by alpha contributions. This feature has been labeled in panel (g) and is visible in the other VDFs (c-l). In contrast, to the solar wind in Figure \ref{fig3}, the energy is lower, as expected due to the low speed and temperature. As expected from Figure \ref{fig4}, there was no strong anisotropy and the VDFs did not experience significant evolution across this interval to account for the strong MM activity. Surprisingly, the VDFs I and II were similar, which could suggest the change in $|\mathbf{B}|$ and $\beta_i$ was not responsible for sufficiently altering the MM stability. This is reasonable considering that the plasma appeared MM unstable or marginally stable (according to the presence of isolated MM structures) before the sharp onset of these waves. Thus, open questions were raised about this event and there was no immediate local driving mechanism. It could be that this MM criterion does not include key factors that were important to the growth rate of these waves \cite{pokhotelov2002}. On the other hand, it could be that these waves were convected from a different location. These will be addressed later in the discussion.

\subsection{Event 3: 2021-08-14}
\begin{figure}
\noindent\includegraphics[width=0.75\textwidth]{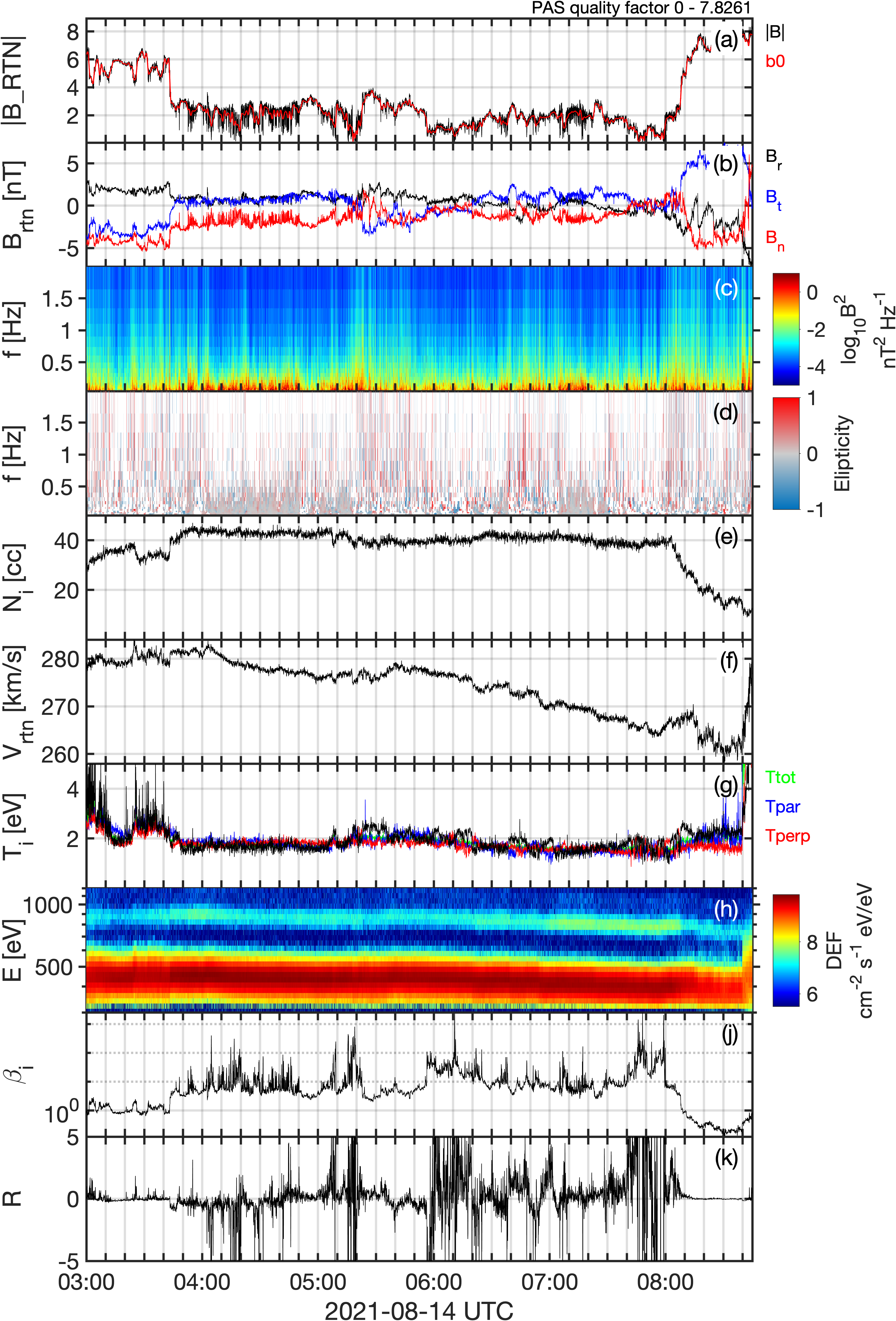}
\caption{Mirror modes observed on 2021-08-14. Plotted in panels (a \& b) are $|\mathbf{B}|$ and $B_{rtn}$, a wavelet spectrogram of $|\mathbf{B}|$ is shown in panel (c), and the ellipticity of the magnetic field is shown in panel (d). Panels (e-k) depict $N_i$, $|\mathbf{V_i}|$, $T_i$, DEF, $\beta_i$, and $RMM$, respectively.}
\label{fig7}
\end{figure}

On 2021-08-14, SolO observed another interval of prolonged MM activity lasting approximately 50 minutes at 0.69 AU from the Sun. These measurements are shown in Figure \ref{fig7} and the panels are organized in the same manner as Figures \ref{fig2} and \ref{fig4}.

Remarkably comparable to the other events, the solar wind speed was still unusually low ($< 300$ kms$^{-1}$). This event offered striking similarities to Figure \ref{fig2} where the MMs appeared in a magnetic depression, high density ($\sim 43$ cm$^{-3}$), and enhanced $\beta_i$. The magnetic field spectral power up to 1 Hz was also visibly intensified. Within the magnetic dip, there were negligible variations of the plasma parameters, but the magnetic field was varying significantly, causing the $\beta_i$ to fluctuate and consequently result in large changes in $RMM$. The ion VDF was also moderately gyrotropic. What was also meaningful regarding this particular event was that the timescales appeared larger than the previous event ($\sim 13$ sec) and therefore $L_{mm} \sim 56 \rho_p$. In addition, the distance between the dips had grown and one can recognize individual MM structures even on this larger time scale. In addition, the amplitudes exceed 1 nT, which is larger than the aforementioned cases. 

After scrutinizing multiple cases of MM storms, there seemed to be two distinguishable types of events. Type one corresponds to intensive bursts with timescales around 1 second and amplitude up to 1 nT, which manifested as peaks or dips and had spatial scales of around one or several ion gyroradii. Type two was consistent with extended trains of magnetic holes and had timescales of several seconds, amplitudes larger than 1 nT, and larger spatial scales that were several 10s of the local gyroradii. The features of this event could also suggest another HCS encounter similar to Figure \ref{fig2}. Thus it seemed HCS crossings were effective at setting up MM growth conditions.

\subsection{Short-term temporal evolution of mirror mode structures}
So far, the case studies that were presented revealed MM intervals in which the individual structures were invariant in many properties, such as peakness, frequency, and the spacing between peak/dip structures. Here, the peakness refers to if the MMs were peaks or dips and was determined from the skewness of the probability distribution of magnetic field calculated from:
 \begin{linenomath*}
\begin{equation}
S = \frac{M_3}{\sigma^3},
\end{equation}
 \end{linenomath*}
where
 \begin{linenomath*}
\begin{equation}
M_3 = \frac{1}{N}\sum^{N}_{i=1}(B_i-\bar{B})^3.
\end{equation}
 \end{linenomath*}
For $S<0$, the MMs were dips and when $S>0$ the MMs were peaks. In this section, examples are shown that demonstrate the evolution of peakness and other MM properties.

\begin{figure}
\noindent\includegraphics[width=0.5\textwidth]{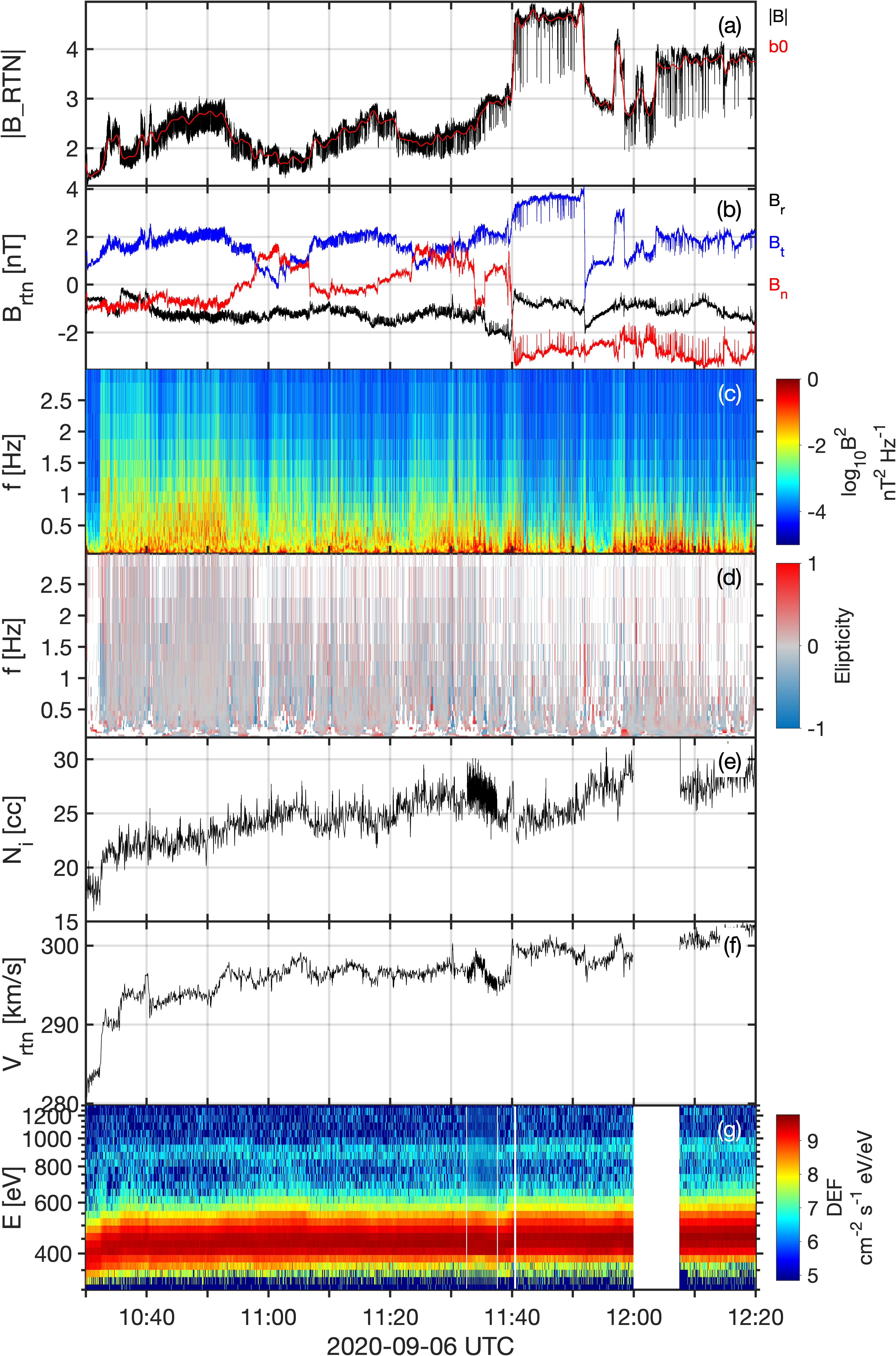}
\caption{MMs on 2020-09-06. Panels (a \& b) show the magnetic field, plotted in panels (c \& d) are a wavelet transform of $|\mathbf{B}|$ and ellipticity, respectively. The remaining panels (e-g) are $N_i$, $|\mathbf{V_i}|$, and DEF.}
\label{fig8}
\end{figure}

Throughout this investigation, two types of MM intervals were discussed. These types were defined based on the frequency and amplitude of the structures. Yet, it is necessary to point out that these different types were not mutually exclusive nor did they have to occur within completely separate events. Depicted in Figure \ref{fig8} is a period of intense MM activity on 2020-09-06. Panels (a, b, and c) depict the magnetic field time series and a wavelet transform of its magnitude. The ellipticity is plotted in panel (d) where the prolonged linear polarization is easily identified by the nearly zero ellipticity. The remaining panels (e-g) show $N_i$, $V_i$, and the DEF. The temperature was reliable enough to draw conclusions from, so it was not included. Note that there were some data gaps in the plasma measurements resulting in absent data in the bottom three panels, which does not interfere with the investigation.
Comparable to the previous cases, the event took place during a slow solar wind stream with speeds lower than 300 kms$^{-1}$ and high densities ($N>20$ cm$^{-3}$). Between 10:40 and 11:40, the MMs were around 1 Hz with amplitudes around 0.5 nT. At 11:40 there was a polarity reversal of $B_n$ and a significant change in $B_r$. Following this, there was a prominent change in the MM structures resulting in larger amplitudes ($>1$ nT), larger periods, and increased proximity between individual structures. Thus, from 11:30 - 11:40, the physical nature of the MMs had dramatically changed, which seems triggered by the field rotation and a slight increase in $|\mathbf{V_i}|$.

\begin{figure}
 \noindent\includegraphics[width=0.5\textwidth]{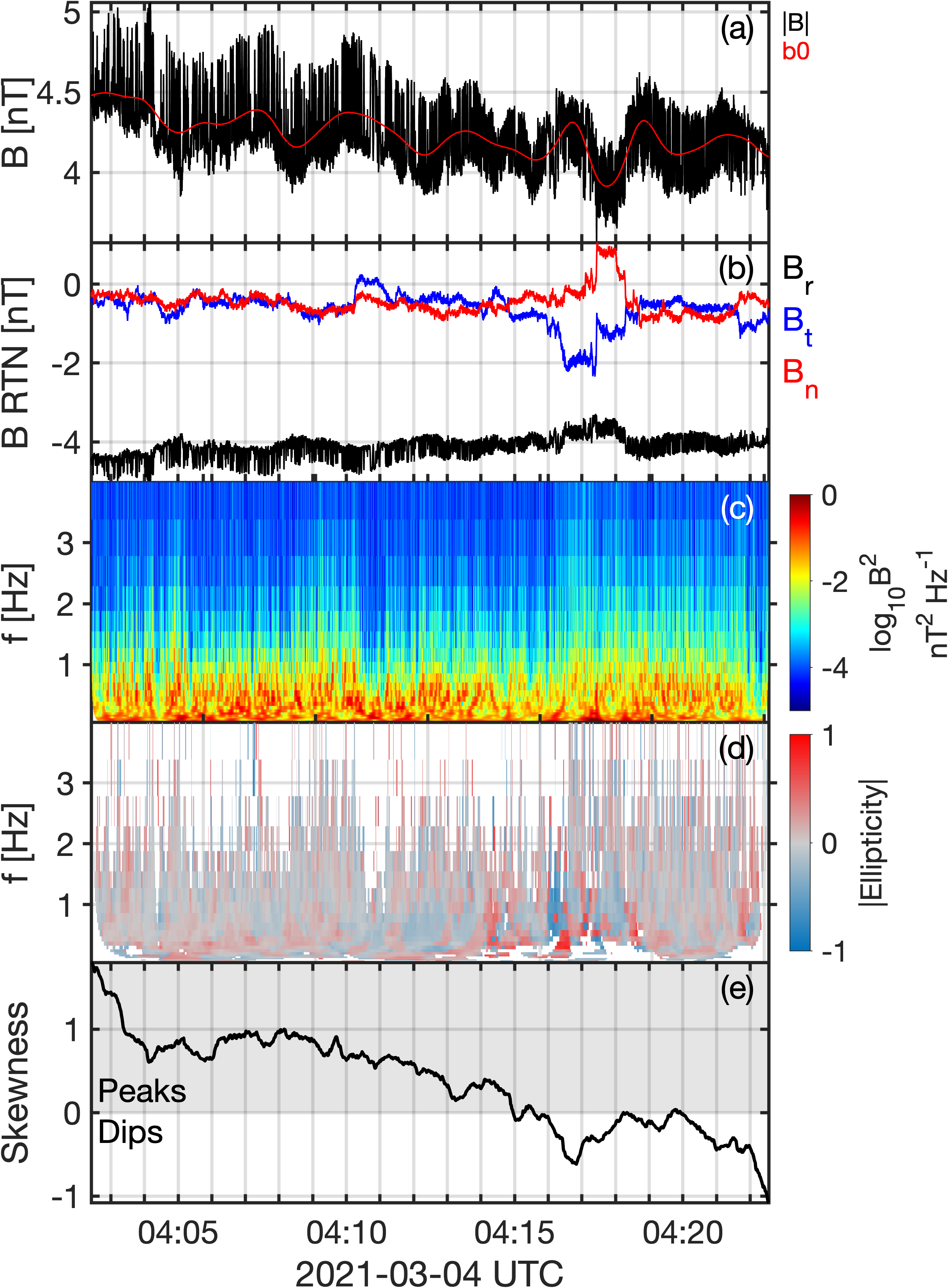}
\caption{The evolution of MMs observed on 2021-03-04. The top two panels (a \& b) show the mirror mode structures in the magnetic field whereas a wavelet of $|\mathbf{B}|$ is shown in panel (c). The ellipticity of the magnetic field is shown in panel (d) and the skewness below in panel (e). The skewness ($S$) demonstrates whether the mirror modes are peaks (skewness $S>0$) or dips ($S<0$) and panel (e) indicates a change in $S$ over this interval.}
\label{fig9}
\end{figure}

In addition to the evolution of amplitude and frequency as shown in Figure \ref{fig8}, the peakness could similarly deviate. In the next example, this occurred over approximately 10 minutes. Figure \ref{fig9} shows a case where the MMs evolved from peaks to dips. It is interesting to note that there was an interval with circularly polarized waves, however, the analysis of such waves was not within the scope of the present study. Panels (a, b, and \& c) correspond to the magnetic field time series and a wavelet transform of the magnetic field. Plotted in panel (d) is the ellipticity whereas panel (e) is the skewness calculated over a sliding window of 20 seconds that was advanced by one second until the end of the interval was reached.
The skewness slowly transitions from positive to negative, indicating a shift from peaks to dips, respectively. Yet, the frequency seemed to remain constant throughout. Thus, there was no sharp change in conditions responsible, contrary to the previous example; the change occurred more gradually. But, there was a small rotation in $\mathbf{B}$ around 04:16, however, the peakness evolution seems to be underway prior to this. Unfortunately, particle measurements were not available during this time, so it is not possible to interpret the plasma conditions. Based on earlier studies \cite{soucek2008,genot2009,soucek2015,dimmock2015}, MM peaks are associated with more MM unstable (larger value of $RMM$) conditions. Hence, this evolution may signify a transition from MM unstable to marginally MM stable conditions.

\subsection{Mirror mode storm downstream of an IP shock by MMS}
\begin{figure}
\noindent\includegraphics[width=0.75\textwidth]{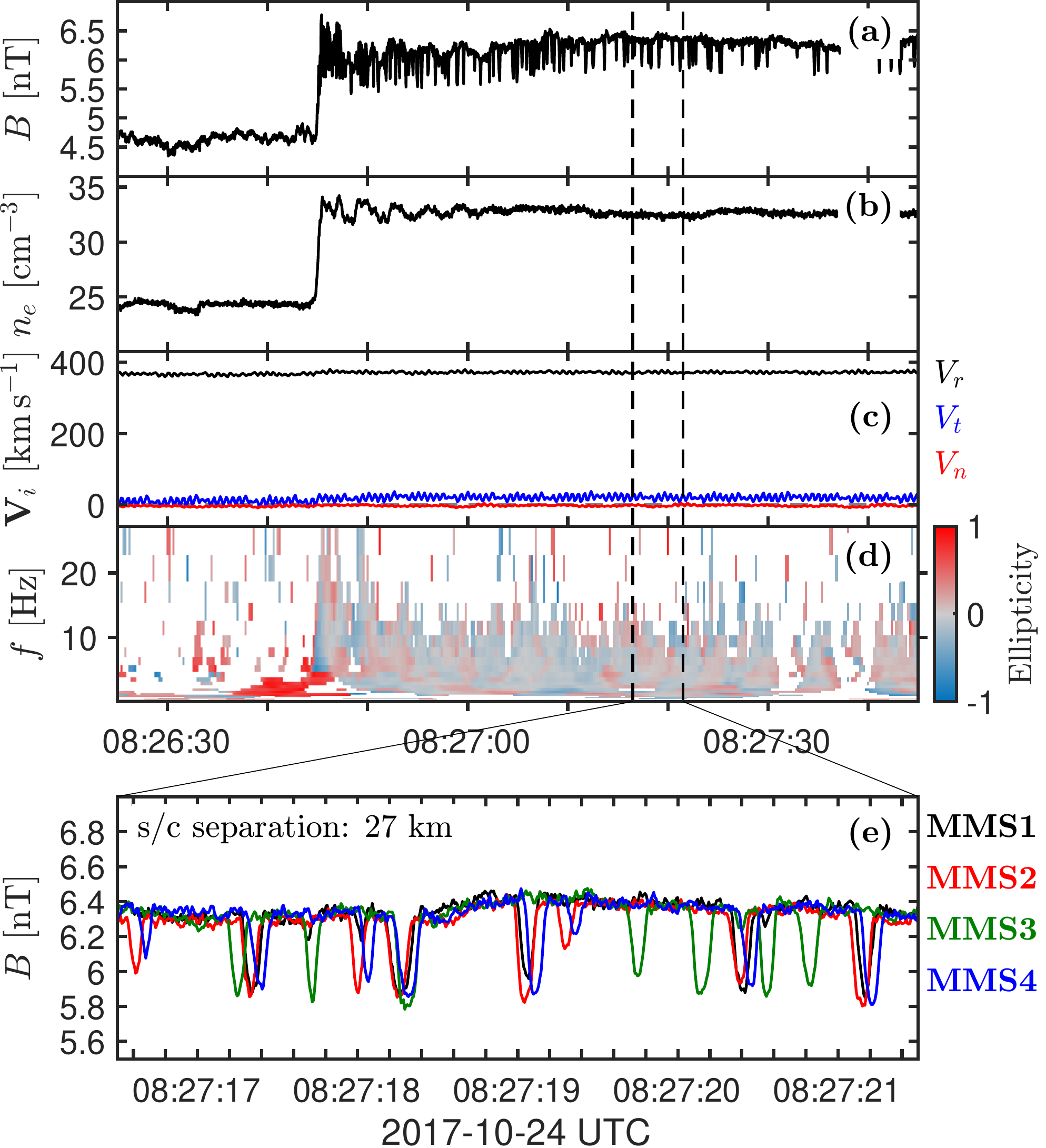}
\caption{A MM storm observed by MMS downstream of an IP shock on 2017-10-24. Panels (a-d) show $|\mathbf{B}|$, $n_e$, $|\mathbf{V_i}|$ and the polarization of $\mathbf{B}$, respectively. The bottom panel displays a shorter interval where all four MMS spacecraft are plotted together.}
\label{mms_IP_shock}
\end{figure}

Since SolO is a single spacecraft, the assumption of zero propagation in the plasma rest frame has been used to make conclusions regarding the spatial scale of the MMs. However, MMS consists of 4 spacecraft with inter-spacecraft separations that are sometimes similar to the spatial scale of the MMs that were studied with SolO. Thus, MMS can be used to directly infer the spatial scales. For this reason, this section describes MMS observations of MMs that were observed directly downstream of an IP shock on 2017-10-24. This is shown in Figure \ref{mms_IP_shock}, and although IP shocks have been reported \cite{cohen2019, hanson2020}, this event is likely to be the earliest known IP shock measured by MMS. Panels (a-d) show $|\mathbf{B}|$, $n_e$, $\mathbf{V_i}$, and ellipticity. The bottom panel is a zoomed-in plot of $|B|$ but shows all four MMS spacecraft. In this specific example, $n_e$ is a plasma moment as opposed to derived from the spacecraft potential, which was the case with SolO. The shock crossing was oblique ($\theta_{bn} \sim 52^\circ$) and low Alfv\'en Mach number ($M_A \sim 1.7$).
Almost immediately downstream from the shock ramp, there is a sudden onset of a MM train. Only burst mode is shown here but the MM structures can be observed for around 3.5 minutes after the shock ramp. The structures are linearly polarized and appear as sharp dips, similar to some of the other events studied with SolO. What is valuable in this example is that the multi-point measurement can be used to directly infer the spatial scales of the individual structures. What is interesting here is that in panel (e), some MM structures are observed by some MMS spacecraft, but not by others. This implies that these structures are on the same scale, or smaller than the spacecraft separation. Here, the average spacecraft separation is 27 km and $\rho_p \sim 43$ km, confirming that these MMs are smaller than the local ion scales. This is true in at least one direction, however, the 3D geometry of MMs was not assessed here. In addition, the magnetic pressure is balanced by the electron thermal pressure inside the dips. Thus, it is likely that these were electron or kinetic MMs, which will be discussed in more detail below.

\section{Statistical results}
Using SolO, it is now feasible to investigate events ranging over heliocentric distances ($|\mathbf R|$) without the reliance upon separate spacecraft conjunctions. In addition, the onboard suite of instruments allowed the investigation into the solar wind conditions that are key to the growth of MMs. Regardless, a manual search is laborious and impractical. Therefore, an automated search was employed. As established by these case studies and prior literature, clear characteristics of these events where $\mathbf{B}$ had a high degree of polarization ($>0.8$),  linearly polarized (ellipticity = 0), anti-correlated with density, and manifested as trains of structures continuing for several minutes. It is also important to reiterate that the purpose was to identify train-like MM events and not isolated magnetic holes. Having stated that, the amplitudes of these structures could be low ($<$ 1 nT) and although they were visible in the magnetic field, this was not always the case for the plasma density. For this rationale, the lengthy period of a high degree of polarisation concurrent with low ellipticity was used. It is also necessary to mention that utilizing the local plasma conditions such as temperature anisotropy may have been helpful in this search. However, these measurements were not available for lengthy periods, whereas the magnetic field is consistently available. In addition, it appears that the existence of MMs does not always correspond with the anticipated in situ plasma conditions such as $T_{\perp i}>T_{\parallel i}$. For these reasons, the automated search was performed using solely magnetic field measurements.

\subsection{Automated search}
The search was conducted on measurements between 2020-04-15 and 2021-08-31. The step-by-step procedure was as follows:
\begin{enumerate}
\item Compute the magnetic field ellipticity ($\epsilon$) and degree of polarization within a 5-minute window between frequencies 0.1-2 Hz (0.5s-10s).
\item Apply a mask to points where the degree of polarization falls below 0.7.
\item Require that 75\% of $|\epsilon| < 0.2$.
\item Save the times of windows that satisfy the criteria of predominantly being linearly polarized.
\item Advance the window by 2.5 minutes (50\% overlap) and repeat.
\end{enumerate}
The above process delivered a set of 5-minute windows that fulfilled these criteria. These windows were then manually arranged into separate events and visually inspected for signatures of MM structures. If events were separated by more than 1 hour, these were documented as separate intervals. The outcome was 25 separate intervals, which are listed in table \ref{tab_events} as well as some essential parameters.
 \begin{sidewaystable}
 \caption{Prolonged MM events observed with Solar Orbiter detected by the automated search. Listed are some fundamental properties as well as the type of structure that the MMs were associated with. The acronyms SW-CS and SBC refer to solar wind current sheet and sector boundary crossing, respectively. }
 \label{tab_events}
 \centering
 \begin{tabular}{l l l l l l l l l l c c}
 \hline
\# & Date [UTC] & Type & $|\mathbf{V_i}|$ [kms$^{-1}$] & $N_i$ [cm$^{-3}$] & Ti [eV] & T$_{\perp/\parallel}$ & $\beta_i$ & Event Type & QF & $|R|$ [AU] & CDF ver [MAG, PAS]\\
 \hline
 	1  & 2020-04-16 12:00 & 2 & & & & & &   & & 0.82 &[1, n/a] \\
	2  & 2020-04-16 22:56 & 2 & & & & & &   & & 0.82 &[1, n/a] \\
	3  & 2020-06-04 08:30 & 2 & & & & & &   & & 0.54 &[3, n/a] \\
	4  & 2020-07-08 11:38 & 2 & 311 $\pm 1 $ & 18 $\pm 1$ & 5 $\pm 0.4$ & 0.90 $\pm 0.28$ & 5.2 $\pm 5.1$ & SW-CS & 0.26 $\pm 0.15$ & 0.60 &[3, 3] \\
	5  & 2020-07-29 10:33 & 1 & 297 $\pm 1 $ & 13 $\pm 1$ & 3 $\pm 0.2$ & 0.54 $\pm 0.09$ & 4.5 $\pm 1.3$ &  SW-CS & 7.98 $\pm 1.49$ & 0.73 &[3, 3] \\
	6  & 2020-08-09 11:02 & 1 & 326 $\pm 3 $ & 23 $\pm 1$ & 3 $\pm 0.1$ & 0.97 $\pm 0.33$ & 3.5 $\pm 1.8$ & SW-CS & 0.04 $\pm 0.06$ & 0.80 &[2, 3] \\
	7  & 2020-08-27 02:55 & 1 & & & & & &  & & 0.89 &[2, n/a] \\
	8  & 2020-09-06 10:32 & 1, 2 & 297 $\pm 3$ & 26 $\pm 3$ & 2 $\pm 0.5$ & 0.90 $\pm 0.24$ & 3.2 $\pm 1.3$ & SIR & 0.59 $\pm 0.71$ & 0.93 &[2, 3] \\
	9  & 2020-09-06 16:31 & 2 & 336 $\pm 3$ & 45 $\pm 2$ & 3 $\pm 0.3$ & 1.07 $\pm 0.09$ & 3.1 $\pm 1.1$ & SIR & 0.00 $\pm 0.0$ & 0.93 &[2, 3] \\
	10 & 2020-12-02 12:20 & 2 & & & & &  & & & 0.87 &[4 ,n/a] \\
	11 & 2020-12-31 21:40 & 1 & & & & & & & & 0.70 &[4 ,n/a] \\
	12 & 2021-01-27 18:58 & 2 & & & & & & & & 0.53 &[4 ,n/a] \\
	13 & 2021-01-30 04:40 & 2 & & & & & & & & 0.52 &[3 ,n/a] \\
	14 & 2021-02-20 12:44 & 1 & & & & &  & & & 0.51 &[2 ,n/a] \\
	15 & 2021-03-04 04:00 & 1 & & & & & & & & 0.57 &[2 ,n/a] \\
	16 & 2021-04-07 16:14 & 1 & & & & & & & & 0.79 &[2 ,n/a] \\
	17 & 2021-04-19 13:55 & 1 & & & & &  & & & 0.85 &[3 ,n/a] \\
	18 & 2021-05-31 08:05 & 1 & 321 $\pm 5$ & 16 $\pm 1$ & 3 $\pm 0.2$ & 0.92 $\pm 0.1$ & 1.5 $\pm 0.3$ & SW-CS & 0.00 $\pm 0.00$ & 0.95 &[2, 3] \\
	19 & 2021-05-31 10:24 & 1 & 319 $\pm 3$ & 31 $\pm 3$ & 7 $\pm 2.8$ & 0.51 $\pm 0.17$ & 4.7 $\pm 1.8$ & SW-CS & 0.10 $\pm 0.10$ & 0.95 &[2, 3] \\
	20 & 2021-06-27 08:46 & 1 & & & & & &  & & 0.92 &[1, n/a] \\
	21 & 2021-06-28 03:42 & 1, 2 & & & & & &  & & 0.92 &[1 ,n/a] \\
	22 & 2021-07-19 10:18 & 1 & 427 $\pm 9$ & 61 $\pm 14$ & 12 $\pm 2.4$ & 1.17 $\pm 0.24$ & 1.5 $\pm 2.1$ & SIR \& SBC & 0.00 $\pm 0.00$ & 0.84 &[1, 3] \\
	23 & 2021-08-14 04:10 & 2 & 278 $\pm 1$ & 43 $\pm 1$ & 2 $\pm 0.1$& 1.00 $\pm 0.09$ & 10.9 $\pm 20.6$ &  SBC& 0.31 $\pm 0.12$ & 0.69 &[2, 3] \\
	24 & 2021-08-14 07:05 & 2 & 270 $\pm 1$& 41 $\pm 1$ & 2 $\pm 0.1$& 1.02 $\pm 0.08$ & 10.3 $\pm 15.8$ &  SBC& 0.65 $\pm 0.19$ & 0.69 &[2, 3] \\
	25 & 2021-08-14 18:57 & 2 & 286 $\pm 1$& 33 $\pm 1$ & 4 $\pm 0.3$ & 1.54 $\pm 0.13$ & 2.8 $\pm 2.0$ &  SBC& 0.23 $\pm 0.09$ & 0.69 &[2, 3] \\
	 \hline
\multicolumn{8}{l}{$^{}$$\pm$ calculated as one standard deviation}
 \end{tabular}
 \end{sidewaystable}

The quantities documented in table \ref{tab_events} represent the mean values over the interval that MM structures were visually perceptible. As demonstrated by the case studies above, this can be a variable period between several minutes to an hour. Thus, quantities can deviate significantly. For this reason, this variability has been denoted by adding $\pm$ one standard deviation. The PAS quality factor (QF) has also been included to provide readers with a proxy for the trustworthiness of these values.
 
From table \ref{tab_events}, there are several intriguing results to point out. Surprisingly, all of the events (when PAS data was available) were identified during moderate/slow solar wind streams. This could also be a statistical effect caused by the average speed of the solar wind in the studied period, which will be discussed later. However, an anomaly is the SIR on 2021-07-19 ($|\mathbf V_i| \sim 430$ kms$^{-1}$) compared to the remaining events where $|\mathbf V_i|<350$ kms$^{-1}$. As anticipated from the slow solar wind, $N_i$ was also high and as depicted in the case study shown above that surpassed 80 cm$^{-3}$. For most events, $T_i$ was low, but the temperature anisotropy could be highly complex and variable during each event. To some extent, this explains why the anisotropy values in table \ref{tab_events} are generally moderate and why in some cases $T_{\perp i}<T_{\parallel i}$. A caveat to interpreting these values properly is that this cannot be considered the ambient solar wind, that is, solar wind that is not clearly associated with some known transient such as an SIR and ICME. Even so, the occurrence during slow solar wind is striking.

\subsection{Dependence on heliocentric distance and solar wind conditions}
\begin{figure}
\noindent\includegraphics[width=0.4\textwidth]{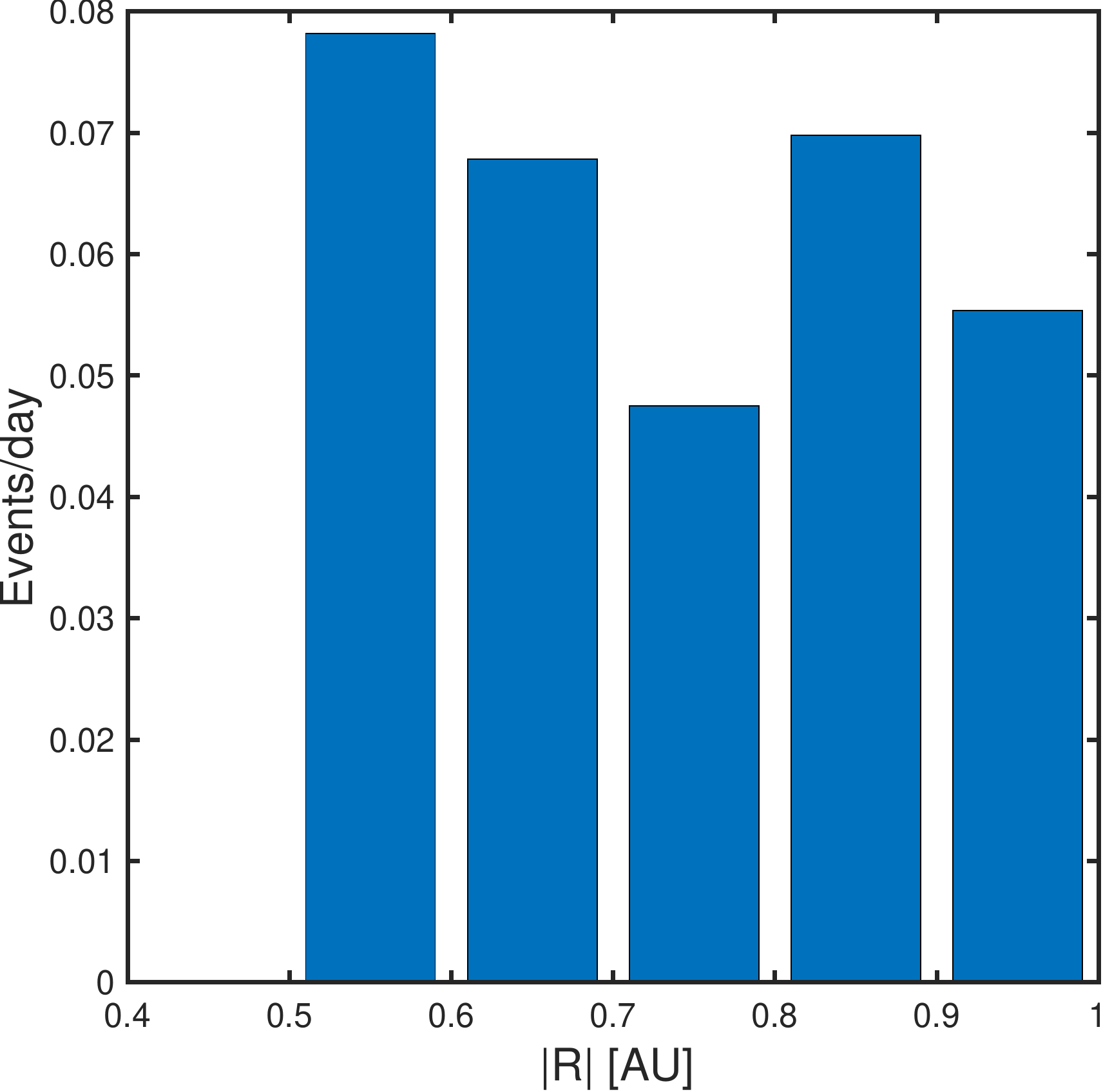}
\caption{Occurrence rate of prolonged mirror mode trains across heliocentric distances between 0.5-1 AU.}
\label{fig10}
\end{figure}

The criteria adopted in this automated search were intended to identify prolonged intervals of linearly polarised structures that are indicative of MMs. From a period of 16.5 months, only 25 intervals were detected. Although further events would be desired to more accurately calculate the occurrence rates of these events, one prominent result was that their presence is not frequent. It was also possible to calculate the probability at which $|\mathbf R|$ these events are identified. This is plotted in Figure \ref{fig10}.
The values in Figure \ref{fig10} are calculated based on the availability of $|\mathbf B|$ such that event counts were normalized by the availability of MAG data at each $|\mathbf R|$ bin. Although the number of events is limited to 25, Figure \ref{fig10} implies that the likelihood of identifying these events declines with raising $|\mathbf R|$. Having said that, this trend was not strong, and additional events will be required for confirmation.

\begin{figure}
\noindent\includegraphics[width=\textwidth]{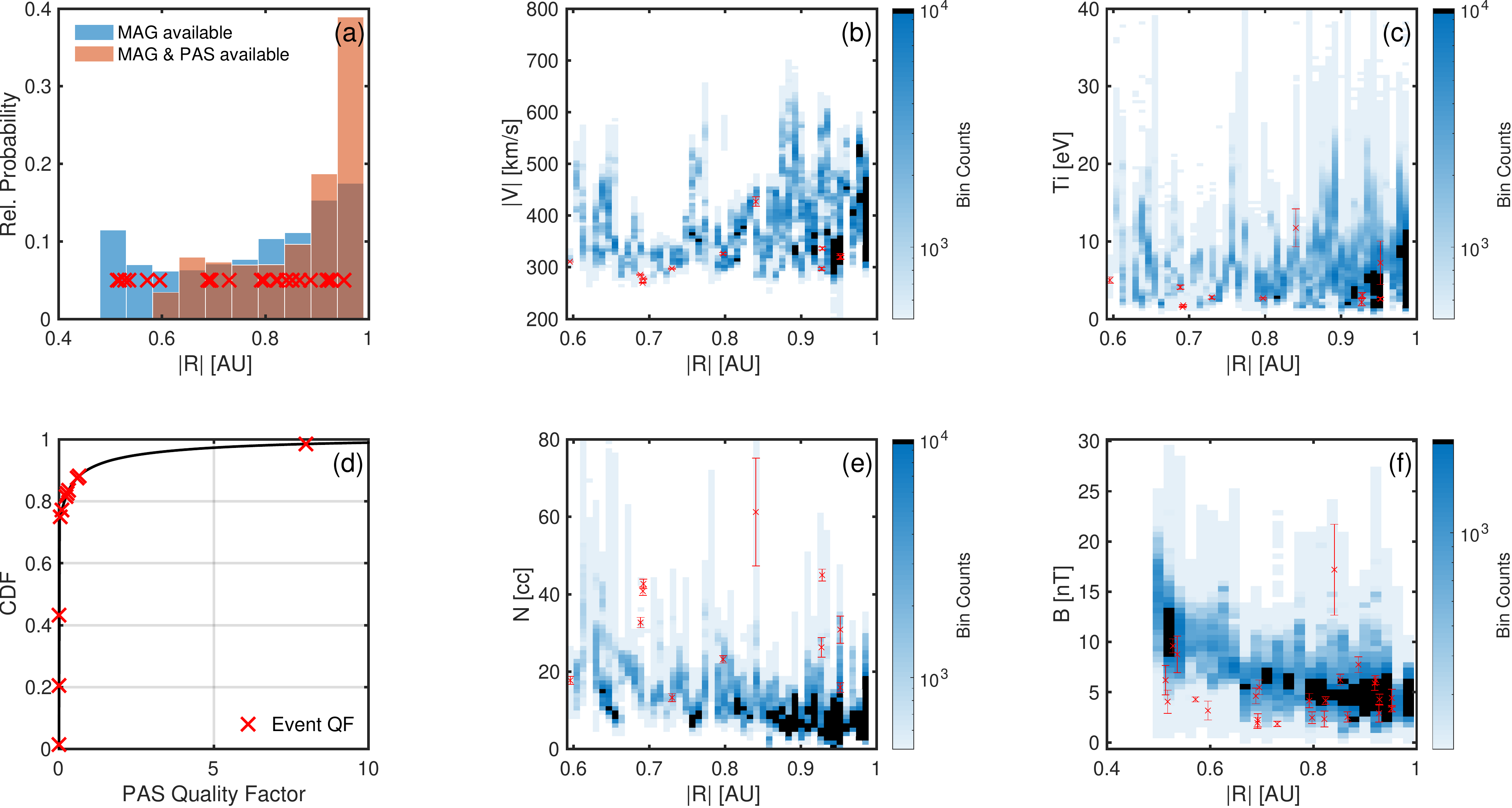}
\caption{Solar wind statistics measured by SolO between $0.5<R<1$. Panel (a) shows the availability of data when MAG and MAG+PAS data are available. Panels (b, c, e, \& f) show $|\mathbf{V_i}|$, $N_i$, and $|\mathbf{B}|$ as a function of $|R|$ in which the color shows the bin counts and the red crosses are the values for specific events. Panel (d) is the cumulative distribution function for the PAS quality factor.}
\label{fig11}
\end{figure}

To properly interpret the values provided in table \ref{tab_events}, they have to be put into context with the typical values of the solar wind. However, these will vary with $|\mathbf R|$, which was investigated in Figure \ref{fig11}. 
Panel (a) shows the availability of MAG and MAG+PAS data for bins of $|\mathbf R|$. Thus, the spacecraft occupied $|\mathbf R| \sim 1$ longer than $|\mathbf R| \sim 0.5$. This demonstrated why this had to be taken into account in the occurrence rates of these events. The red crosses show the values of $|\mathbf R|$ for each of the 25 events (note the placement on the y-axis is arbitrary). Plotted in panels (b, c, e, and f) are 3D histograms of various quantities for bins of $|\mathbf R|$. The 3D histograms are used here to demonstrate the availability of solar wind data at various heliospheric distances. The red crosses again show the values for each event and the error bars correspond to $\pm$ one standard deviation. The cumulative distribution function (CDF) of the PAS quality factor is located in panel (d).

According to panel (b), and as expected, the solar wind speed naturally increased with $|\mathbf R|$ \cite{Khabarova2018}, however in general the MM events stayed at the lower range of $|\mathbf V|$ regardless of $|\mathbf R|$. Thus, based on the criteria that were adopted here, the events were identified within the slow solar wind for each heliocentric distance. However, there is a lack of faster solar wind speed observations at some $|\mathbf R|$, particularly around 0.7 AU and 0.85 AU, and the bin density corresponding to slower speed is higher at 1 AU. Thus, it cannot be ruled out completely that there could be some statistical influence. Panel (c) also demonstrated that events also occurred during cold ion temperatures, and according to panel (e), higher than typical ion densities. This could be related to the fundamental characteristics of the fast and slow solar wind, i.e. the slow solar wind is usually denser and colder. However, it should be noted that these events were not identified in the ambient solar wind. There was a tendency for $|\mathbf B|$ to decrease with $|\mathbf R|$, but there was no clear reliance on the magnetic field strength of the events depicted in panel (f). The case studies presented in detail above were selected partly based on low-quality factor values from PAS (i.e. high-quality data), however, panel (d) suggested that several events suffer from higher quality factors which are unavoidable due to the low solar wind speed for each event. It should also be noted that many events appeared as outliers for ion density and for a couple of cases in the magnetic field. This is expected since these MM storms did not seem to appear in the ambient solar wind but during disturbed intervals such as SIRs and current sheets.

\section{Discussion}
For the first time, missions such as SolO and Parker Solar Probe (PSP) have enabled the study of the dependence of kinetic instabilities and other complex structures such as MMs on heliocentric distance ($<$ 1 AU) and solar wind properties. This paper has concentrated on continuous MMs, referred to in prior studies as mirror mode storms \cite{russell2009,enriquez_rivera2013} that differ from the more isolated magnetic hole structures that are examined in numerous earlier studies \cite{turner1977, Winterhalter1994, Xiao2010, volwerk2020, karlsson2021}. The objective here was to understand the connection with the solar wind, structures/transients, heliocentric distance, and local plasma conditions while shedding light on their physical properties. Throughout this investigation, several case studies were analyzed followed by statistical results. Statistics were compiled utilizing an automated search exploiting the linearly polarized nature of these types of structures. There were multiple physical mechanisms/structures over a wide variety of temporal scales that created the conditions favorable for MM growth. Yet, although significant questions remain for some cases, several clear and novel conclusions could be reached. Below, the physical interpretations of these results are discussed, and explanations for difficult events are offered, which are also put into context with the existing literature.

According to the MM growth condition (equation \ref{eq:R}), the larger the $\beta_i$, the smaller the temperature anisotropy needed for the plasma to become mirror unstable \cite{hasegawa1969, soucek2008}. In two events studied, HCS crossings resulted in simultaneous magnetic field decreases and density increases \cite{Simunac2012}, which created sudden and large enhancements of $\beta_i$. Such conditions should then require only small temperature anisotropies to set up MM unstable plasma conditions. It seems HCSs can sometimes be embedded within SIRs and CMEs, and one result obtained from this study demonstrated that they were highly efficient at setting up conditions for MM growth. This also implied that the plasma parameters (e.g. temperature anisotropy and $\beta_i$) associated with large-scale solar wind transients such as SIRs and CMEs were also constrained to some extent by these instabilities. Similar to planetary magnetosheaths \cite{soucek2015,Genot08,dimmock2015}, solar wind transients also offer a rich natural laboratory for investigating these structures.

For the events when $\beta_i$ was low, the temperature anisotropy did not appear to reach exceptionally large values and appeared constrained between 1 and 1.2, sometimes even below 1. These events, therefore, appear marginally stable or near the stability threshold; some were noticeably below. At Earth, and planetary magnetosheaths in general, MMs are mainly driven by the large temperature anisotropy created by the quasi-perpendicular bow shock \cite{dimmock2013,dimmock2015,soucek2015,osmane2015}, which also increases the $\beta_i$. Interplanetary shocks also produce temperature anisotropies, which can result in mirror modes \cite{ala-lahti2018}. However, in that study, they appeared as more isolated magnetic hole structures and not as the MM storms that were examined here. Two interplanetary shocks occurred for the event presented in Figures \ref{fig1} and \ref{fig2} and both shocks did appear to generate moderate temperature anisotropy downstream. However, the increase in $\beta_i$ is not significant since the density and magnetic field both increase across the shocks, and the ion temperature change is inconsequential. This was evident from panel (c) in Figure \ref{fig1} as no sharp changes in $\beta_i$ occurred across the shock fronts. It seemed that the interplanetary shocks studied here do not seem to be efficient in generating the conditions resulting in MM storms. On the other hand, \citeA{russell2009} confirmed that MM storms can be generated downstream of weak interplanetary shocks, so this is not always the case. No MM storms were observed directly downstream from SolO shocks in this study but were driven by large-scale changes in field and plasma properties associated with other structures. However, one cannot rule out shorter MM intervals that fall outside of the search criteria adopted here. Analyzing other SolO interplanetary shocks (not shown) also implied that MM storms are not a common feature. \citeA{enriquez_rivera2013}, also proposed that shocks were not essential to MM storm growth in their investigation, which used STEREO data. Nevertheless, this could be highly specific to shock parameters (e.g. Mach number, geometry) and it is worthy of more research as SolO assembles a diverse shock catalog over the nominal mission phase and beyond. 

\citeA{enriquez_rivera2013} also reported that the alpha particle density increased for most of the MM storms that they studied. Although alpha particle moments were not directly available for this study, there was some evidence to support enhanced alpha particle density in some events (e.g. Figures \ref{fig1} and \ref{fig7}) during the enhanced $\beta_i$ intervals. It has been established in previous studies \cite{price1986,hellinger2005,lee2017} that different particle species can play a significant role in modifying the mirror mode instability criteria while also having the effect of suppressing the competing ion cyclotron instability. However, it was not possible to directly investigate that in this study since the instrument does not separate ions and alphas.

It is worth commenting that some of the events identified in this study showed no evident local mechanisms for MM growth, particularly because the ion temperature anisotropy was around one  (e.g. Figure \ref{fig4}) while $\beta_i$ was also small. There are some conceivable explanations for these events. Firstly, the events identified in this study occurred during low solar wind speeds, which can lead to instrumental problems. Explicitly, this can result in nonphysical ion VDF features due to low solar wind energies. This issue is quantified to some capacity by the PAS quality factor, which serves as a proxy that is anti-correlated to the trustworthiness of the data. As a rule-of-thumb, the PAS quality factor increases for low solar wind speeds, and the data becomes less reliable. In addition, the temperature is a higher-order moment and is especially susceptible to artifacts in the ion VDFs. As a result, estimating the correct temperature anisotropy becomes challenging in specific situations. Secondly, in cases when the data is reliable, the in-situ plasma measurements may not reflect the MM growth conditions at the moment/location that the structures were generated. The reason was that MMs are convected with the plasma flow, therefore, it is conceivable that the source region could be located elsewhere. Another interpretation is a temporal variation of the source region plasma parameters, such as a relaxation of the temperature anisotropy as a result of the MMs. The final reason stems from the variety of these events in terms of the plasma conditions, spatial scales, and their presence in different solar wind transients. Therefore, a growth condition that incorporates additional factors (e.g electron temperatures, smaller wavelengths, non-Maxwellian VDFs, and other particle species) may be required. 

\begin{figure}
\noindent\includegraphics[width=0.5\textwidth]{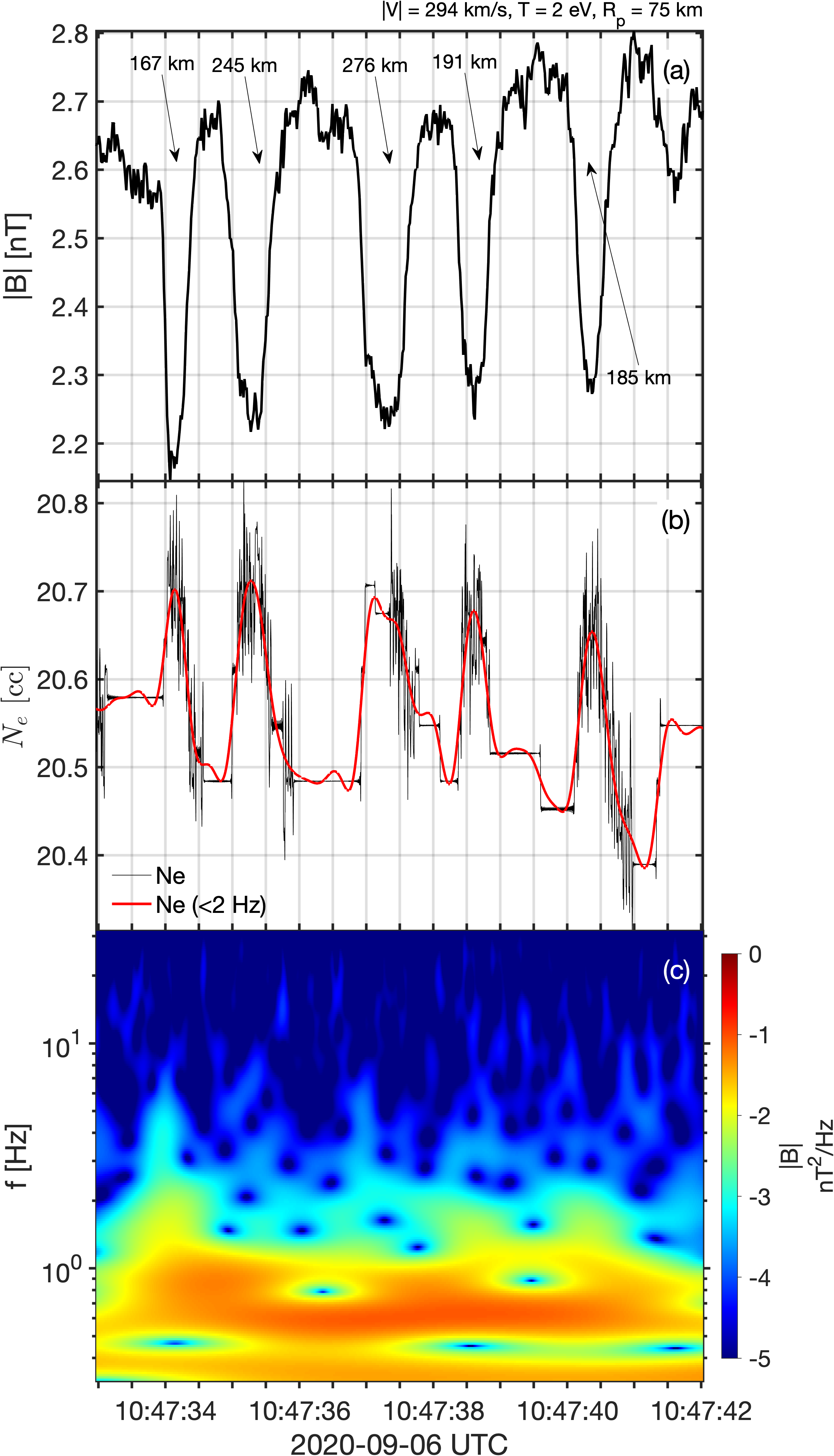}
\caption{Shortened interval from the event on 2021-08-14. Panels (a-c) show $|B|$, $N_e$, and a wavelet of $\mathbf{B}$. The anti-correlation between $|B|$ and $N_e$ is clear when viewed on this timescale.}
\label{fig13}
\end{figure}

The mirror instability threshold expressed in equation \ref{eq:R} \cite{hasegawa1969} is a cold electron bi-Maxwellian fluid approximation, assuming the low frequency and long wavelength limit such that $\omega \ll \omega_{ci}$, $\omega \ll k_\parallel v_A$, and $k_\parallel/k_\perp \ll 1$. This set a quasi-MHD constraint on the spatial scales, meaning MMs were required to be much larger than the local ion scales. In the terrestrial magnetosheath, mirror mode spatial scales are typically a few thousand km (15 sec duration with 150-200 kms$^{-1}$ plasma flow) \cite{soucek2008}, which results in scales at least an order of magnitude beyond the usual ion gyroradii. For most of the events studied here, this condition seemed appropriate, however, some MM structures approached this limit. For example, plotted in Figure \ref{fig13} are several individual MM structures over approximately 10 seconds. Panels (a-c) portrayed $|\mathbf{B}|$, $N_e$ and a wavelet transform of $\mathbf{B}$. Note in panel (b), for clarity, the red trace indicated a 2Hz low-pass filter of $N_e$. As expected, $N_e$ was anti-correlated with $|\mathbf{B}|$.
By measuring the duration of each structure, the spatial scale could be estimated from the plasma flow since MMs have zero phase speed in the plasma rest frame. During this interval, $\rho_p \sim 75$ km and $L_{mm} = 167-276$ km, hence these MM structures were approaching the ion kinetic scales. Although the unusually slow solar wind raised the quality factor and reduced the reliability of the data, it is plausible to consider the solar wind speed was slow as it is expected from the other events. To confirm and strengthen this result, a MM train was found in MMS data. Using multiple spacecraft, directly confirms that these MM trains can be smaller than local ion scales. Moreover, even in this case, the solar wind speed was $< 400$ kms$^{-1}$, which is consistent with the SolO events. As a result, these cases may test the low-frequency limit assumption and a fully kinetic MM threshold may be demanded.

For cases when $L_{mm} \gg \rho_p$, it has been indicated that finite electron temperature effects in the long-wavelength limit also modify the instability threshold \cite{pantellini1995,pokhotelov2002}. This occurs due to the electron pressure gradient that in turn generates an $E_\parallel$ \cite{pantellini1995}, increasing the mirror mode instability threshold and lowering the growth rate. Nevertheless, it is not anticipated that this would shed meaningful light on the ambiguous events reported here since the ion anisotropy was weak for these cases; which, would only contribute to explaining a lack of MMs during large anisotropies. Another key consideration was that during CMEs and SIRs, the particle distributions are expected to deviate from the non-bi-Maxwellian shape due to the existence of characteristics such as shocks, sheaths, and current sheets. Prior work \cite{pokhotelov2002} had sought to address this by understanding the consequences of arbitrary distribution functions (within the long-wavelength limit). The consensus from that study was that distributions such as loss cones and tails from energetic particles can reduce the instability threshold and increase the growth rate. On the other hand, although the VDFs examined here did present slight deviations from non-Maxwellianity, there was no evidence of significant features such as energetic particles and/or supra-thermal tails. Although these effects cannot be ruled out entirely, it was not expected to play a considerable role here; but they may become more consequential in explaining MM growth in additional plasma regimes or solar wind transients. Note that PAS VDFs are collected every 4 seconds, so the purpose was to check the underlying distribution properties in relation to the occurrence of the MM trains. For larger-scale MMs lasting several seconds, it could be possible to investigate the ion dynamics and particle interactions with MM structures (e.g. \citeA{soucek2011}). 

When $L_{mm} \sim \rho_p$ or below, the electron-scale mirror mode threshold $RMM_{e} = (T_{e\perp}/T_{e\parallel}) / (1 + 1/\beta_{e\perp})>1$ \cite{pokhotelov2013} can explain the generation of MMs. This was shown experimentally by \citeA{yao2019} who studied such structures upstream of the Earth's bow shock using MMS. The authors showed that even though there was no ion temperature anisotropy, the presence of an electron temperature anisotropy was understood to provide the sufficient free energy required. But it should be pointed out that the structures analyzed in that study were 0.1 $\rho_p$, which is smaller than the approximately $2 \rho_p$ that were calculated for the SolO cases here. Although it should be noted that the MMS case presented in Figure \ref{mms_IP_shock} was significantly less than the $\rho_p$. Electron MMs do not apply to all the events in this investigation, but only in cases when the spatial scales approach or are below ion scales. One feature consistent with the cases in this study was the clear anti-correlation with $N_e$. Nevertheless, an investigation into the physics of kinetic MM structures is outside the scope of this study, but it should be considered for future investigations of these structures using electron data.

The statistical analysis has revealed several intriguing results. Firstly, all but one of the 25 events were found when $|\mathbf{V}|<400$ kms$^{-1}$. A straightforward explanation is that the median solar wind speed for the data set that was analyzed was 340 kms$^{-1}$, thus the probability of finding events for $|\mathbf{V}|<400$ kms$^{-1}$ was not so unreasonable. Thus, one explanation could be statistical. Yet, this does not justify the lack of event detection when the solar wind speed was faster, since there were data available according to Figure \ref{fig11}, especially at 1 AU. Some solutions could be discovered from the various studies that have devoted efforts to understanding the radial evolution of solar wind parameters (e.g. \citeA{Khabarova2018,echer2020}), and the inter-dependency of properties during fast and slow solar wind streams. Yet, this is not readily applied to the current study and will not be explored further here. The reason is that MM storms did not tend to appear in the ambient solar wind, but were associated with transients such as SIRs, HCS, and other field and plasma structures. Hence, the ambient solar wind properties could be misleading in this regard as they are more applicable to isolated MMs such as magnetic holes, which are abundantly found in the ambient solar wind. Thus, this remains an open question, but as SolO collects more data into the following solar cycle, forthcoming studies will shed more light on this. These results also imply that the probability of detecting MM storms is higher closer to the Sun. This could be an indication of the tendency for events to occur for lower solar wind speeds ($|\mathbf{V_i}|$ increases with radial distance), but the same arguments above are valid and it is problematic to apply undisturbed solar wind conditions. Thus, future studies could concentrate on the evolution of solar wind transients and determine if ``younger" SIRs and/or CMEs are more prone to these instabilities. Opportune radial alignments (e.g. SolO, PSP, BepiColombo, ACE/Wind) may also shed light on this topic. The final point to make is that MM storms are not common. Just 25 events were identified between 15 April 2020 and 31 August 2021. The broader implications of this propose that MM storms should not play a meaningful role in regulating and/or constraining the ambient solar wind properties. On the other hand, MM storms should be more crucial to solar wind transients and complicated structures, primarily during high $\beta_i$ conditions.

This study also showed that MM trains can also undergo significant deviations in terms of their amplitudes and frequency. This was especially pronounced in two examples that were highlighted (see Figures \ref{fig8} and \ref{fig9}). According to earlier studies \cite{soucek2008,Genot08,dimmock2015}, peaks are associated with MM unstable plasma ($RMM>1$), whereas dips tend to occur for marginally stable MM conditions ($0<RMM<1$); but are able to survive the transition to MM stable plasma. Thus, the interpretation of these events is that the change in peakness (peaks-dips) is owed to local changes in plasma conditions that deviate to more marginally MM stable conditions. The change from peaks to dips was also noted by \citeA{enriquez_rivera2013}. In Figure \ref{fig8} the temporal width of the MMs increased from 0.7 seconds to 3.2 seconds across the event even though the plasma speed remained stable. The time between individual structures also increased from $<$1 second to $>$1 minute. In the immediate vicinity, there is a reversal in $B_{n,t}$, an increase of $|\mathbf{B}|$, and a decrease in $N_i$. Thus, $\beta_i$ decreases, which could push the plasma to more mirror stable conditions, explaining the change from peaks-dips. The difference in frequency is also connected to the above discussion, where the initial spatial scales are $\sim 2 \rho_p$, implying other factors may need to be assessed. Thus, the change of frequency, in this case, could demonstrate an evolution of electron temperature and/or the move toward satisfying the long-wavelength limit assumption. \citeA{russell2009} postulated that MM storms may evolve as they are carried outward by the solar wind, as they similarly behave in the magnetosheath when moving towards the magnetopause. Although deviations in properties seem to take place for individual events as debated above, there was no clear evidence yet to point towards a fundamental discrepancy between the properties of these waves at smaller heliocentric distances compared to those at 1 AU. An important caveat to consider in this work is the criteria for the automated search, which analyzed 5 minutes windows. Thus, the search could have missed shorter interval MM trains that were notably shorter than the window length.

The present study has achieved its goals by shedding significant light on the properties of MM storms in the solar wind, their dependence/occurrence with heliocentric distance, and their connection to large-scale transients. The study has also highlighted the complex nature of MM storm and their occurrence across a wide variety of plasma structures. Naturally, some open questions remain, especially when MMs violate the long-wavelength assumption and the mechanisms responsible for their growth are unclear. With increasing catalogs of inner-heliospheric observations from SolO, PSP, and BepiColombo, these data are, and will, be a rich source for advancing understanding of the coupling between kinetic instabilities and large-scale structures. In addition, closer-than-before perigees ($\sim$ 0.3 AU) will provide new insights into where and when such instabilities develop and the importance of the ``age" of solar wind transients.

\section{Summary \& conclusions}
The objective of this study was to shed important light on continuous mirror mode activity in the solar wind, previously called mirror mode storms. The main motivation was the scarcity of literature on the topic, which the Solar Orbiter mission is ideally placed to fill. The study has utilized Solar Orbiter data from 2020-04-15 - 2021-08-31 between heliocentric distances of 0.5-1 AU, resulting in 25 events. Several events were studied in detail whereas some statistical analysis was presented later. From this work, the main conclusions can be summarized as follows:
\begin{enumerate}
\item A statistical search based on magnetic field data only detected MM storms during moderate-slow solar wind speeds.
\item Heliospheric current sheet, interplanetary current sheets, and extended magnetic field minima appear to be efficient at setting up conditions for MM growth due to sudden enhancements of $\beta_i$.
\item MM storms manifest over a range of spatial scales, but in some situations approach the local ion gyroradii, which challenges the long-wavelength limit assumption.
\item Based on the events considered here, interplanetary shocks were not the dominant driver of MM storms. However, with increasing solar activity this could change as more shocks are expected.
\item MM storms demonstrate visible evolution in terms of peakness, spatial scale, and amplitude. \item MM storms typically arise in two categories, the first has a higher frequency (1-2 Hz) and smaller amplitudes ($<$1 nT) and can appear as peaks. The second has amplitudes $>$1 nT and frequencies $<$ 1 Hz and seems to appear as dips.
\item The typical temporal scales of individual MMs are between 0.5 - 1.5 seconds, but this can be larger.
\item MM storms are not common, and only 25 events were detected between 2020-04-15 and 2021-08-31.
\item Due to the low occurrence, MM storms likely do not play a major role in modifying the ambient solar wind properties. However, their importance in terms of regulating the plasma should increase during large-scale disturbed intervals such as SIRs and CMEs.
\item There is evidence to suggest that MM storms are more likely to be observed at smaller heliocentric distances between 0.5-1 AU. However, more events will be required to provide a definitive confirmation.
\item For some events, it was not clear what plasma conditions were responsible. One interpretation was that finite electron temperatures, kinetic scales, and non-Maxwellian distribution functions need to be accounted for. Or it could be that the MMs were generated elsewhere. Another likely possibility was that the alpha particle population may play a strong role. However, currently, there are no readily available alpha particle moments to properly assess their role, which could be addressed in a future study.

\end{enumerate}

\acknowledgments
APD received financial support from the Swedish National Space Agency (Grant 2020-00111) and the EU Horizon 2020 project SHARP: SHocks: structure, AcceleRation, dissiPation 101004131. 

Solar Orbiter is a space mission of international collaboration between ESA and NASA, operated by ESA. Solar Orbiter Solar Wind Analyser (SWA) data are derived from scientific sensors which have been designed and created, and are operated under funding provided in numerous contracts from the UK Space Agency (UKSA), the UK Science and Technology Facilities Council (STFC), the Agenzia Spaziale Italiana (ASI), the Centre National d’Etudes Spatiales (CNES, France), the Centre National de la Recherche Scientifique (CNRS, France), the Czech contribution to the ESA PRODEX programme and NASA. Solar Orbiter SWA work at UCL/MSSL is currently funded under STFC grants ST/T001356/1 and ST/S000240/1. We thank the entire MMS team and instrument PIs for the access and use of MMS. 

Solar Orbiter data is publicly available at the ESA Solar Orbiter archive (https://soar.esac.esa.int/soar/). MMS data is freely available from the MMS science data center (https://lasp.colorado.edu/mms/sdc/public/). The OMNI data were obtained from the GSFC/SPDF OMNIWeb interface at https://omniweb.gsfc.nasa.gov.


%
%


%
%
%
%
%

\end{document}